\documentclass[preprint,12pt]{elsarticle}

\usepackage{graphicx}
\usepackage{amssymb}

\usepackage{lineno}
\usepackage{comment}
\usepackage{amsmath}
\usepackage{xspace}
\usepackage{xcolor}
\usepackage{caption}
\usepackage{subcaption}
\usepackage{placeins}

\newcommand*{\MATRIX}{\textsc{matrix}\xspace}
\newcommand*{\MCFM}{\textsc{mcfm}\xspace}
\newcommand*{\pt}{\ensuremath{p_{T}}\xspace}
\newcommand*{\ptV}{\ensuremath{p_{T}^{V}}\xspace}

\newcommand*{\MET}{\ensuremath{E_{T}^{\text{miss}}}\xspace}

\newcommand*{\ptgamma}{\ensuremath{p}_{T}^{~\gamma}\xspace}
\newcommand*{\mll}{\ensuremath{M_{\ell\ell}}\xspace}
\newcommand*{\ZZllnunu}{\ensuremath{ZZ\rightarrow \ell^{+}\ell^{-}\nu\bar{\nu}}\xspace}
\newcommand*{\Zg}{\ensuremath{Z\gamma\rightarrow \ell^{+}\ell^{-}\gamma}\xspace}
\newcommand*{\TeV}{\ensuremath{\text{Te\kern -0.1em V}}}
\newcommand*{\GeV}{\ensuremath{\text{Ge\kern -0.1em V}}}

\newcommand*{\as}{\ensuremath{\alpha_s}}

\journal{JHEP}

\begin{document}

\begin{frontmatter}


\title{Estimating $ZZ$ production with $Z\gamma$ events at the LHC:\\ cross-section ratio and uncertainties}

\author[1]{Jorge Sabater Iglesias}
\author[1]{Vincent Goumarre}
\author[2]{Fang-Ying Tsai}
\author[3]{Mangesh Sonawane}
\author[1]{Sarah Heim}
\author[1,4]{Beate Heinemann}

\address[1]{Deutsches Elektronen-Synchrotron (DESY), Hamburg, Germany}
\address[2]{Department of Physics and Astronomy, Stony Brook University, Stony Brook, NY, USA}
\address[3]{Institut f{\"u}r Hochenergiephysik, Wien, Austria}
\address[4]{Physikalisches Institut, Albert-Ludwigs-Universit{\"a}t Freiburg, Freiburg, Germany}

\begin{abstract}
Standard Model $ZZ$ production is an important background for many searches at the LHC, especially in final states with missing transverse momentum. 
In this article, boson substitution is applied to estimate \hbox{\ZZllnunu} yields from \hbox{\Zg}  events. The cross-section ratio of the two processes and its uncertainties are evaluated at NNLO with the \MATRIX generator as a function of the transverse momentum of the substituted boson. Uncertainties due to higher-order QCD corrections, parton distribution functions, photon isolation criteria, and electroweak corrections are evaluated. They depend strongly on the applied event selections and the considered transverse momentum range. For minimal selections, their size is 3$-$4\%, dominated by QCD-related uncertainties for transverse momenta below 500~\GeV\ and by uncertainties due to the factorization of QCD and electroweak corrections at higher transverse momenta. 
\end{abstract}


\end{frontmatter}


\section{Introduction}
\label{intro}
As the ATLAS~\cite{Collaboration_2008} and CMS~\cite{Chatrchyan:2008aa} detectors at the Large Hadron Collider (LHC) at CERN collect more proton-proton collision data,  the sensitivity of many measurements and searches is significantly affected by theoretical and experimental uncertainties on the background predictions. 
Instead of simulating background processes with Monte-Carlo (MC) techniques, estimating the backgrounds from data can help reduce some of the associated uncertainties.

The \ZZllnunu process is an important background in many searches for new physics, e.g. searches for heavy resonances decaying to two $Z$ bosons~\cite{Aaboud:2017rel,Sirunyan:2018qlb,Aad:2020fpj}, the search for associated production of a $Z$ and a Higgs boson, which decays into invisible 
particles~\cite{Aaboud:2017bja,Khachatryan:2016whc} or searches for Supersymmetry in final states with leptons and \MET~\cite{Aaboud:2018ujj,Sirunyan:2017qaj}. 
In this work, we explore if the \hbox{\Zg} process can be used to estimate the normalisation and \ptV-distribution of the \hbox{\ZZllnunu} process. Here, \ptV\ is the transverse momentum of the $Z$ boson decaying into two neutrinos, identified through missing transverse momentum (\MET) in the detector and often used as discriminating observable in searches for invisible particles.

An ideal sample for estimating the \ZZllnunu yields from data would contain $ZZ\to \ell^+\ell^-\ell^+\ell^-$ events, as the production cross-section is identical and differences occur only due to the different branching ratios as well as lepton selection acceptances and efficiencies. However, the $ZZ\to \ell^+\ell^-\ell^+\ell^-$ process suffers from low event rates due to the small branching ratio of the $Z$ boson decay to charged leptons. 
Therefore the $Z\gamma$ process is proposed here, which benefits from larger event rates while its production mechanisms are very similar to the $ZZ$ process. Accounting for the branching ratio difference and the finite identification efficiencies of charged leptons, typically there are about 10 times less identified $ZZ\to \ell^+\ell^-\ell^+\ell^-$ events than  $\ZZllnunu$ events in a given kinematic region. On the other hand,  we expect about twice the $\Zg$ yield compared to \ZZllnunu, as shown later in this article. In the future, e.g. at the High-Luminosity LHC, it might also be advantageous to combine both approaches. 

The suggested boson substitution is inspired by the usage of $\gamma$+jets production for estimating $Z$+jets backgrounds~\cite{Aaboud:2018ujj,Lindert:2017olm}. 
First, a sample of $\Zg$ events is selected in data with the same lepton requirements as used in the respective search. Typical lepton and photon selections are discussed in Section~\ref{sec:eventselection}. The $\ZZllnunu$ yield is then determined from these events by treating the photon as $\MET$ and reweighting the events by the calculated ratio between the \ZZllnunu and \Zg cross-sections as a function of \ptV (see Sections~\ref{sec:XSpredictions} and \ref{sec:XSandratio}). MC simulations are used to correct for the photon reconstruction efficiency and remaining acceptance effects.

In addition to the statistical uncertainties of the $Z\gamma$ control region, which are reduced with the growing data set at the LHC, and residual experimental uncertainties, theoretical uncertainties arise due to remaining differences between the $ZZ$ and $Z\gamma$ processes, leading to imperfect cancellations in the cross-section ratio. 
In Ref.~\cite{Lindert:2017olm}, methods are laid out to estimate these uncertainties in the context of the extrapolation of $\gamma$+jets to $Z$+jets. 
These methods are closely followed here.
Uncertainties due to the finite order of the QCD calculations, uncertainties in the Parton Distribution Functions (PDFs) and the requirement of photon isolation are considered. Their evaluation is discussed in Section~\ref{sec:qcdunc}. In Section~\ref{sec:cuts}, the effect of additional selection criteria on the cross-section ratio and its QCD-related uncertainties is shown.
In addition, higher-order electroweak (EW) corrections can be substantial at high transverse momenta and the uncertainties on the factorization of QCD and EW corrections are included, as presented in Section~\ref{sec:ewkcor}. 

\section{Cross-section predictions for the $ZZ$ and $Z\gamma$ processes}
\label{sec:XSpredictions}
Leading-order (LO) and next-to-leading-order (NLO) cross-section predictions have been available for some time for $ZZ$ \cite{PhysRevD.43.3626,MELE1991409} and $Z\gamma$ \cite{PhysRevD.47.940,PhysRevD.57.2823}.
Recent theoretical developments \cite{Cascioli:2014yka,Grazzini:2015nwa} provide next-to-next-to-leading-order (NNLO) calculations for both processes. 
Higher-order EW corrections have been calculated for diboson processes by various groups~\cite{Hollik:2004tm,Accomando:2005ra,Denner:2015fca,Kallweit:2017khh,Bierweiler:2013dja, Baglio:2013toa,Schonherr:2017qcj,Kallweit:2019zez}. The numbers in this article were provided by the authors of Ref.~\cite{Kallweit:2017khh} based on the OpenLoops generator~\cite{Buccioni:2019sur}.

 Most of the cross-section calculations in this work are performed with \MATRIX~\cite{Grazzini:2017mhc}. The \MCFM event generator~\cite{Campbell:2017aul,Campbell:2011bn}  is used for the evaluation of the PDF and isolation uncertainties.
Proton-proton collisions are simulated at $\sqrt{s}=13$~TeV, using the NNLO CT14~\cite{Dulat:2015mca} PDF set. The results in Section~\ref{sec:cuts} are based on the NNPDF3.0 set~\cite{Ball:2014uwa}. 
The chosen renormalisation and factorisation scales are $\mu_{R,F} = \sqrt{{m_{Z}}^{2} + {\ptV}^{2}}$. 

Diboson production at LO and NLO is quark-initiated: $q\bar{q}\to Z\gamma$/$ZZ$ (including $qg \to Z\gamma$/$ZZ$ at NLO). At NNLO (corresponding to ${\cal O}(\as^2$)), the gluon-initiated production is added: $gg \to Z\gamma$/$ZZ$. This process is often treated separately, e.g. in most ATLAS analyses the $gg\to ZZ$ and $q\bar{q}\to ZZ$ MC predictions are added to obtain a prediction for the total $ZZ$ cross-section~\cite{Aaboud:2017rel,Aad:2020fpj,Aaboud:2017bja}. As the quark-initiated process typically only reaches ${\cal O}(\as$) accuracy in the available (NLO) MC samples, such a combined prediction is not fully consistent in ${\cal O}(\as^2$), but incorporates the leading effect of that higher order.
Very recently, higher-order corrections to the $gg\to ZZ$ process have been calculated~\cite{Grazzini:2018owa}, corresponding to ${\cal O}(\as^3$). 
Since these corrections are not yet available for the $Z\gamma$ process, in this article it is assumed that they are the same as for $ZZ$ production. For the final estimate of the $ZZ/Z\gamma$ cross-section ratio and its uncertainty, these NLO corrections for the $gg$ process are included and the combination of quark-initiated processes at NNLO and gluon-initiated processes at NLO is called nNNLO. 

Collinear divergences arise in the cross-section calculations for QCD radiation at a small angle with respect to the photon, which can be avoided by applying a smooth cone Frixione isolation~\cite{Frixione:1998jh}. 
In this prescription, the photon has to satisfy 
\begin{equation}
\label{eq:frixione_iso}
\sum_{i=\text{partons/hadrons}} p^{i}_{T} (r) \leq \varepsilon_{\gamma}p^{\gamma}_{T} \left( \frac{1-\cos{r}}{1-\cos{R_{0}}} \right)^{n} \hspace{0.5cm} {r\leq R_{0}},
\end{equation}
where $\varepsilon_{\gamma}, R_{0}$ and $n$ are free parameters, which are set in this study to 
\begin{equation*}
    \varepsilon_{\gamma} = 0.075, \hspace{1cm} R_{0} = 0.2, \hspace{1cm}  n = 1.
\end{equation*}  
The smooth cone limits the hadronic activity in the vicinity of the photon, becoming gradually more restrictive closer to the photon.
It was already noted in Ref.~\cite{Lindert:2017olm} that requiring an isolation on the photon can alter the higher-order corrections for $Z\gamma$. This is discussed in~\ref{app:dyncone}.

\section{Event selection}
\label{sec:eventselection}
The baseline event selection in this study follows closely the preselection in a number of ATLAS analyses targeting the presence of a $Z$ boson and missing transverse momentum~\cite{Aaboud:2017rel,Aaboud:2017bja}. Exactly two electrons or muons are required with pseudo-rapidity $|\eta^{\ell}| < 2.5$, and the transverse momentum of the leading (sub-leading) lepton $\pt^{\ell} > 30~\GeV~(20~\GeV)$. The dilepton invariant mass must be in the resonant mass window of the $Z$ boson, $76~\GeV < \mll < 106~\GeV$.  
For the \ZZllnunu process, $E_T^\textrm{miss} > 60$~GeV is required, which corresponds to $\ptgamma>60~\GeV$ for \Zg. 

The photon in \Zg events must be within $|\eta^{\gamma}|<2.5$ and the angular separation between the two leptons and the photon must fulfill $\Delta R(\ell,\gamma) > 0.4$ with $\Delta R = \sqrt{\Delta\phi^2 + \Delta\eta^2}$, where $\phi$ is the azimuthal angle around the beam pipe. This latter cut is primarily applied to reduce the contribution of photons radiating off leptons. 
\begin{table}[h!]
\begin{center}
\begin{tabular}{l | c | c}
\noalign{\smallskip}\hline\noalign{\smallskip}
   Variable & $ZZ$ & $Z\gamma$ \\
 \noalign{\smallskip}\hline\noalign{\smallskip}
 $N_\mathrm{lepton}$ &  \multicolumn{2}{c}{2} \\
  $p_{T}^{\ell_{1}}$ &  \multicolumn{2}{c}{$> 30$ \GeV} \\
  $p_{T}^{\ell_{2}}$ &  \multicolumn{2}{c}{$> 20$ \GeV} \\
$|\eta^{\ell}|$ &  \multicolumn{2}{c}{$< 2.5$ \GeV} \\
  $\mll$ &  \multicolumn{2}{c}{$76$ GeV $< \mll < 106$ \GeV}  \\
  $\MET$ &  $> 60$ \GeV &  $-$ \\    
  $\ptgamma$ & $-$ & $> 60$ \GeV  \\      
  $|\eta^\gamma|$ & $-$ & $<2.5$ \\
  $\Delta R(\ell,\gamma)$ & $-$ & $ > 0.4$ \\
  $\gamma$ isol. & $-$ & Frixione \\
\hline
\end{tabular}
\end{center}
\caption{\label{tab:SRdef} Baseline selection for \ZZllnunu (left column) and $\Zg$ (right column) events.}
\end{table}
Table~\ref{tab:SRdef} gives a summary of the baseline event selection which is used to explore the boson substitution method in this article. 

In many searches for physics beyond the SM, additional selection cuts are placed to further reduce the background while retaining a high efficiency for the signal. 
For instance, in the ATLAS $ZH\to \ell\ell + \MET$ search~\cite{Aaboud:2017bja}, additional selection cuts are placed on observables that depend on the jet activity to suppress spurious \MET from jet-energy misreconstruction. These selections and their impact on the QCD-related uncertainties on the cross-section ratio are described in Section~\ref{sec:cuts}.

\section{$ZZ$ and $Z\gamma$ cross-sections and ratio}
\label{sec:XSandratio}

\begin{figure}[h]
\centering
\begin{subfigure}[b]{0.49\textwidth}
\includegraphics[width=\textwidth]{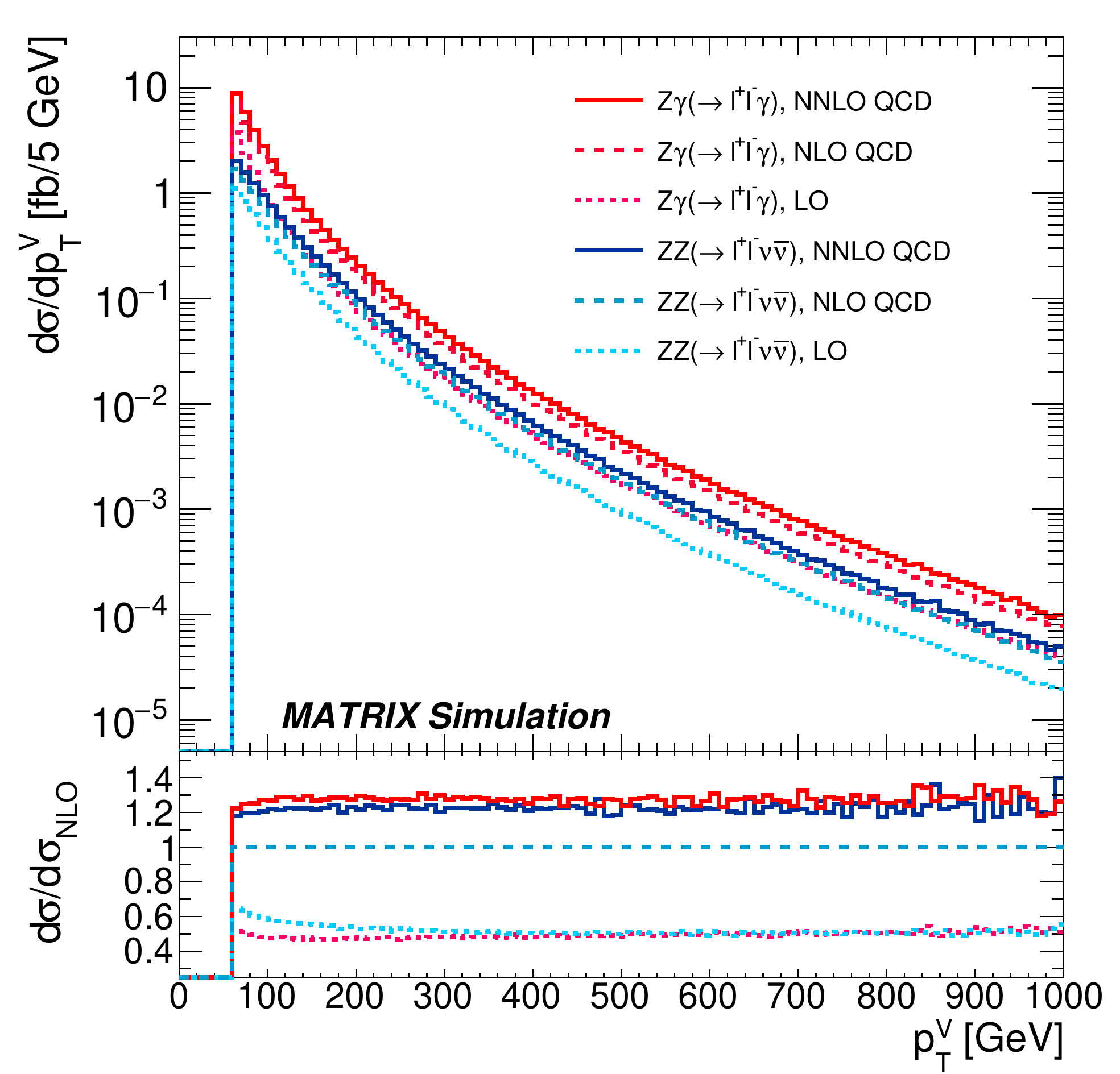}
\caption{}
         \label{fig:xsa}
\end{subfigure}
\hfill
\begin{subfigure}[b]{0.49\textwidth}
\includegraphics[width=\textwidth]{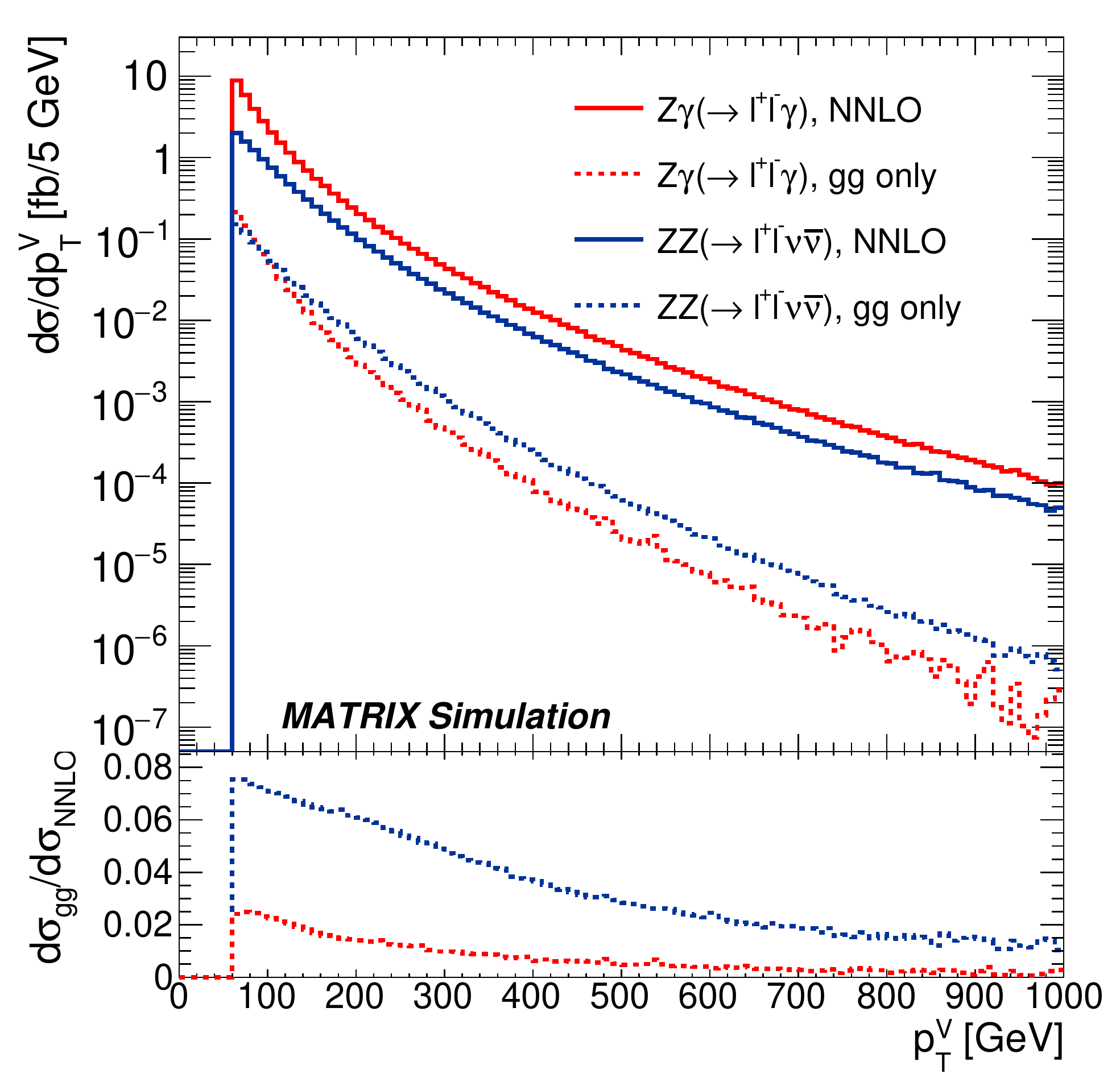}
\caption{}
         \label{fig:xsb}
\end{subfigure}
\caption{(a) $ZZ(\to \ell^+\ell^-\nu\bar{\nu})$ and $Z\gamma(\to \ell^+\ell^-\gamma)$  cross-sections as a function of \ptV at LO (dotted), NLO (dashed) and NNLO (solid) in QCD. The bottom frame shows the ratio of the LO and NNLO calculations to the NLO prediction. (b) \ptV distribution for the cross-sections at NNLO in QCD (solid) and for the $gg$-induced contribution separately (dotted). The bottom frame shows the fractional contribution of the $gg$ process.}
\label{fig:zzzgxs}
\end{figure}

\begin{figure}[h]
\centering
\includegraphics[width=.7\textwidth]{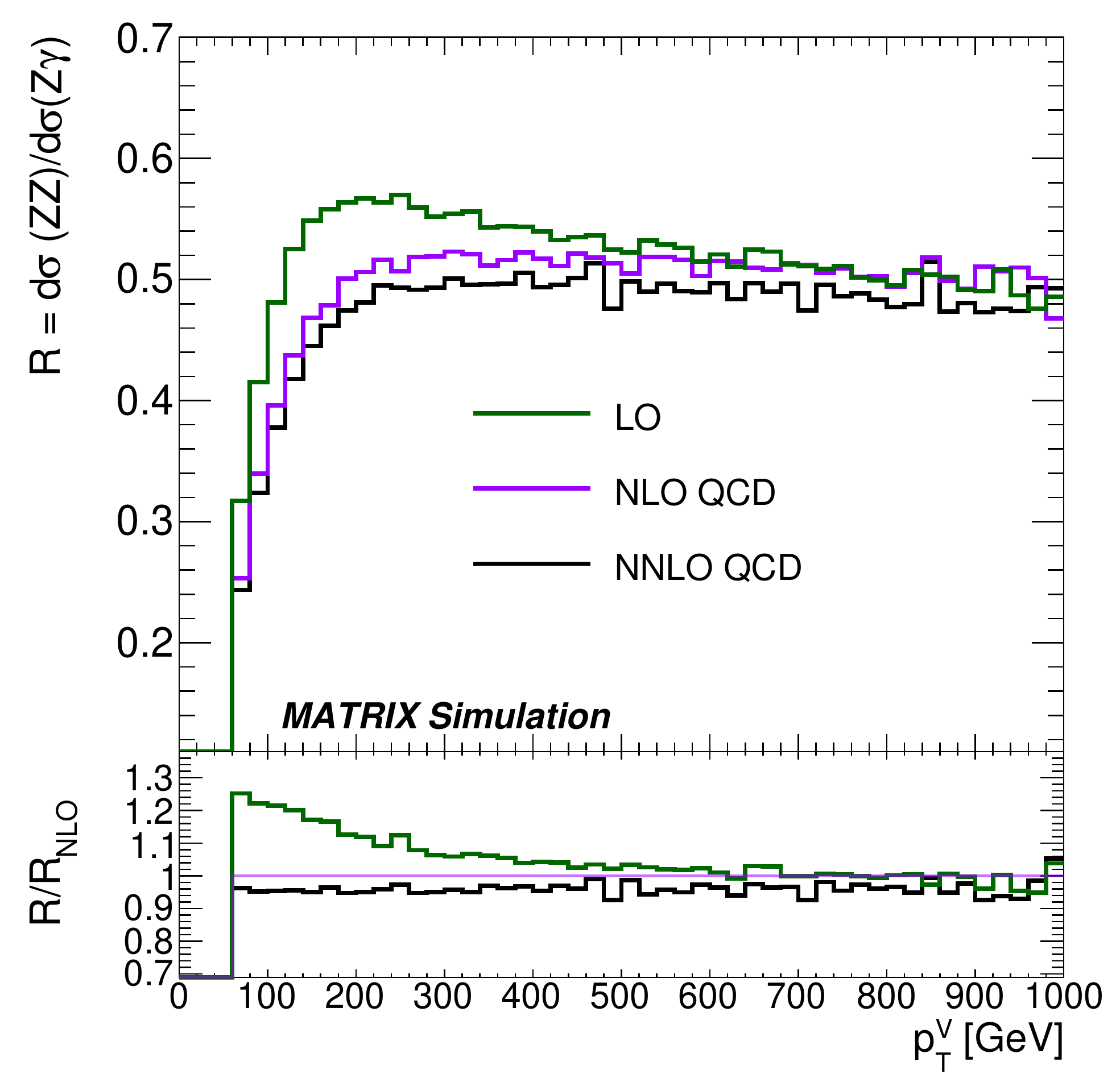} 
\caption{Ratio $R$ between the $ZZ(\to \ell^+\ell^-\nu\bar{\nu})$ and $Z\gamma(\to \ell^+\ell^-\gamma)$ differential cross-sections as function of \ptV at LO (green), NLO (violet) NNLO (black) in QCD. The bottom plot shows the ratios relative to the NLO calculation.}
\label{fig:Ratio}
\end{figure}

The differential cross-sections for $ZZ$ and $Z\gamma$ production as a function of \ptV are shown in Fig.~\ref{fig:zzzgxs} (a) at LO, NLO and NNLO in QCD. The distributions have a very similar shape, particularly for $\ptV>300$~GeV. At lower $\ptV$ there are larger differences. Fig.~\ref{fig:zzzgxs} (b) shows the NNLO cross-sections and the contribution from the $gg$ initial state separately. Here and through this article, unless pointed out specifically by the nNNLO label, the LO $gg$ calculations are used. The $gg$ contribution is largest at low $\ptV$ and is 3$-$4 times larger for $ZZ$ than for $Z\gamma$ production.  
\\
The ratio $R$ of the two cross-section distributions is defined as
\begin{equation}
    \label{eq:ratio}
    R (\ptV) = \frac{d\sigma(\ZZllnunu)/d\ptV}{d\sigma(\Zg)/d\ptV}, 
\end{equation}
and is shown in Fig.~\ref{fig:Ratio} at LO, NLO and NNLO in QCD. It rapidly increases between \ptV values of 60 and 150~GeV and is relatively constant at $R \sim 0.5$ for higher $\ptV$, in particular for the NLO and NNLO calculations. The difference between LO and NLO is most pronounced at low $\ptV$ due to the mass difference between the $Z$ boson and the photon. The ratio of the NNLO to NLO calculation is constant at $\sim$$0.96$.

\section{Uncertainties on the cross-section ratio}
\label{sec:qcdunc}
In this section, the uncertainties on the cross-section ratio $R$ due to QCD aspects of the calculations are discussed, namely higher-order QCD corrections ($\delta^{\rm QCD}$), the photon isolation treatment ($\delta^{\rm iso}$), and PDF ($\delta^{\rm PDF}$). EW corrections and associated uncertainties are discussed in Section~\ref{sec:ewkcor}. 

\subsection{QCD uncertainties}
The applied methodology follows closely the one discussed in Ref.~\cite{Lindert:2017olm}. 
The standard method for assessing uncertainties due to higher (not yet calculated) orders is to vary the renormalisation and factorisation scales by factors of 2 and 0.5. However, it has been shown that this often does not give a complete estimate of the true uncertainty, in particular in cases where a new production mechanism is added at a higher order, such as the $gg$-initiated  process. Since the degree of correlation between the uncertainties on the $Z\gamma$ and $ZZ$ processes is not known, it is also unclear how to vary the scales for the two processes when calculating the ratio $R$. Varying them incoherently likely overestimates the true uncertainty, but a coherent variation, as included below, might be an underestimate.

The first QCD uncertainty component is therefore calculated by checking the convergence of $R$ between two orders. Assuming that the corrections of order $(N+1)$ are smaller than those of order $N$, the size of the missing corrections can be constrained. In this study, the best available calculation is performed at NNLO whereas the size of the next-to-next-to-next-to-leading-order (N$^3$LO) corrections needs to be estimated. 

The $K$-factors on the cross-sections are defined as 
\begin{equation}
    K_{\rm NLO} = \frac{\sigma_{\rm NLO}}{\sigma_{\rm LO}}, 
K_{\rm NNLO} = \frac{\sigma_{\rm NNLO}}{\sigma_{\rm NLO}}, K_{\rm N^3LO} = \frac{\sigma_{\rm N^3LO}}{\sigma_{\rm NNLO}}
\end{equation}
It is clear that identical processes have identical $K$-factors. Turning this around, it should be possible to gauge the similarity of processes by comparing the $K$-factors. 

\begin{figure}[t]
\centering
\begin{subfigure}[b]{0.49\textwidth}
\includegraphics[width=\textwidth]{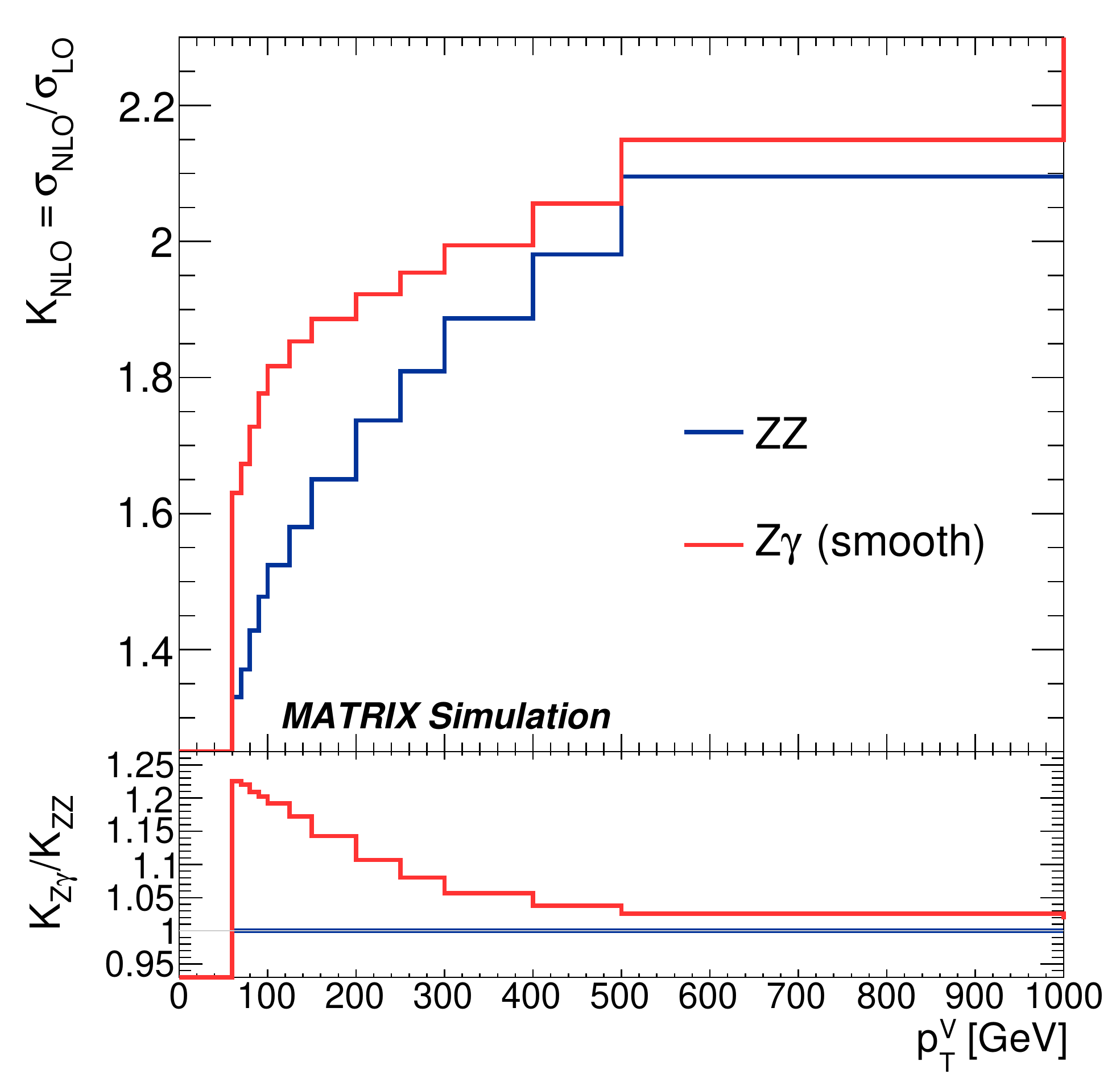}
\caption{}
\end{subfigure}
\hfill
\begin{subfigure}[b]{0.49\textwidth}
\includegraphics[width=\textwidth]{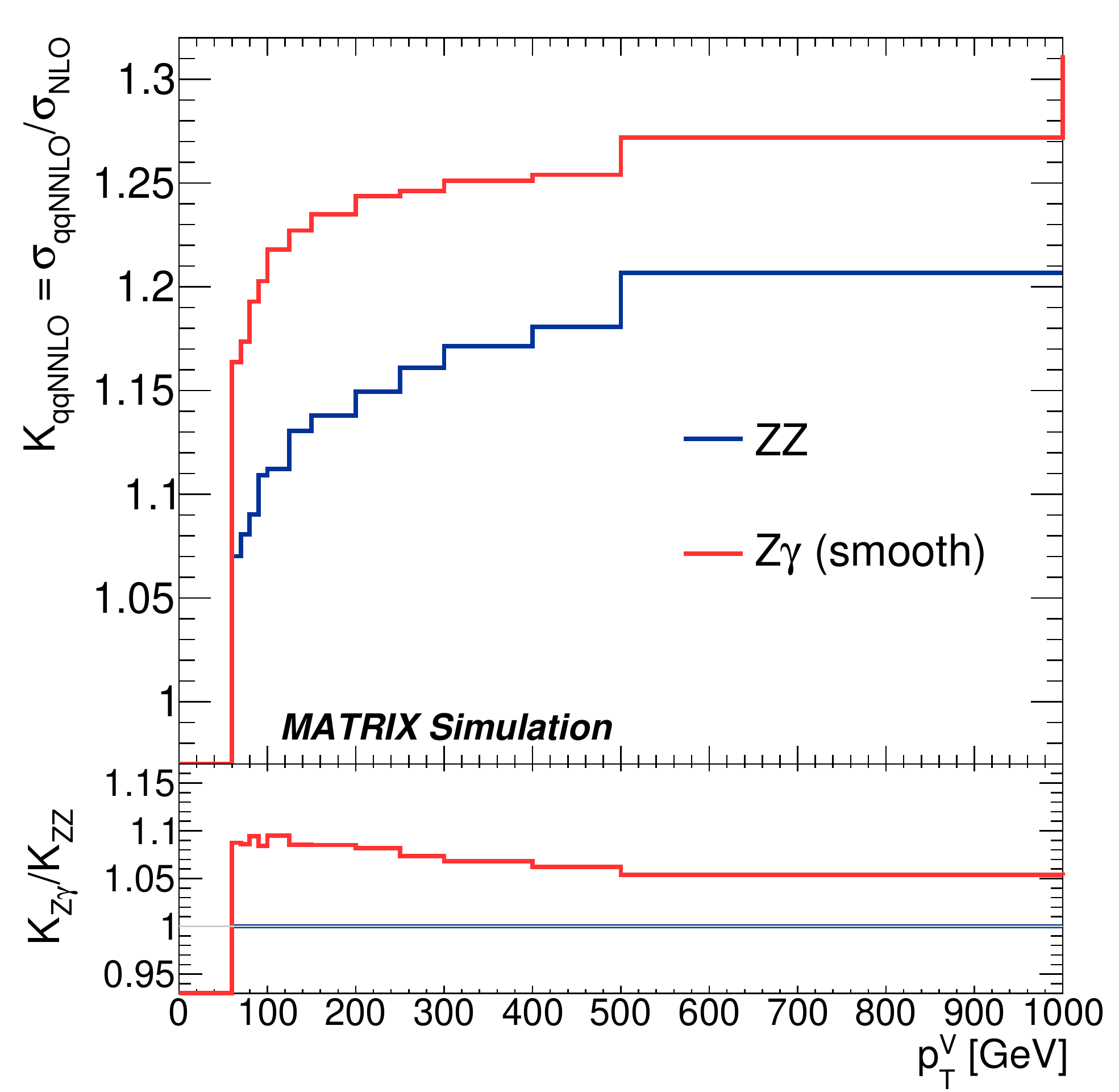} 
\caption{}
\end{subfigure}
\caption{$K_{\rm NLO}$ (a) and $K_{\rm NNLO}$ (b) for $q\bar{q}\to ZZ (\to \ell^+\ell^-\nu\bar{\nu})$ (blue) and $q\bar{q} \to Z\gamma(\to \ell^+\ell^-\gamma)$ with smooth cone isolation (red). The bottom panels show the ratio between the $K$-factors for $Z\gamma$ and $ZZ$ production. 
}
\label{fig:kfactors_nlo_qqnnlo}
\end{figure}

The NLO and NNLO $K$-factors for the $q\bar{q}\to ZZ/Z\gamma$  processes are shown in Fig.~\ref{fig:kfactors_nlo_qqnnlo}. 
It can be seen that for NLO, at low \ptV, the $Z\gamma$ $K$-factors are about 20\% higher than the $ZZ$ $K$-factors, but they approach each other with increasing \ptV. 
The NNLO $K$-factors for the $q\bar{q}\to ZZ/Z\gamma$ processes ($K_\mathrm{qqNNLO}$) have similar trends, but the differences are smaller at low \ptV ($\sim$8\%), decreasing to 5\% at high \ptV.
It is possible to reduce the difference between the $K$-factors further by changing the definition of the isolation cone, as discussed in \ref{app:dyncone}.

\begin{figure}[t]
\centering
\begin{subfigure}[b]{0.49\textwidth}
\includegraphics[width=\textwidth]{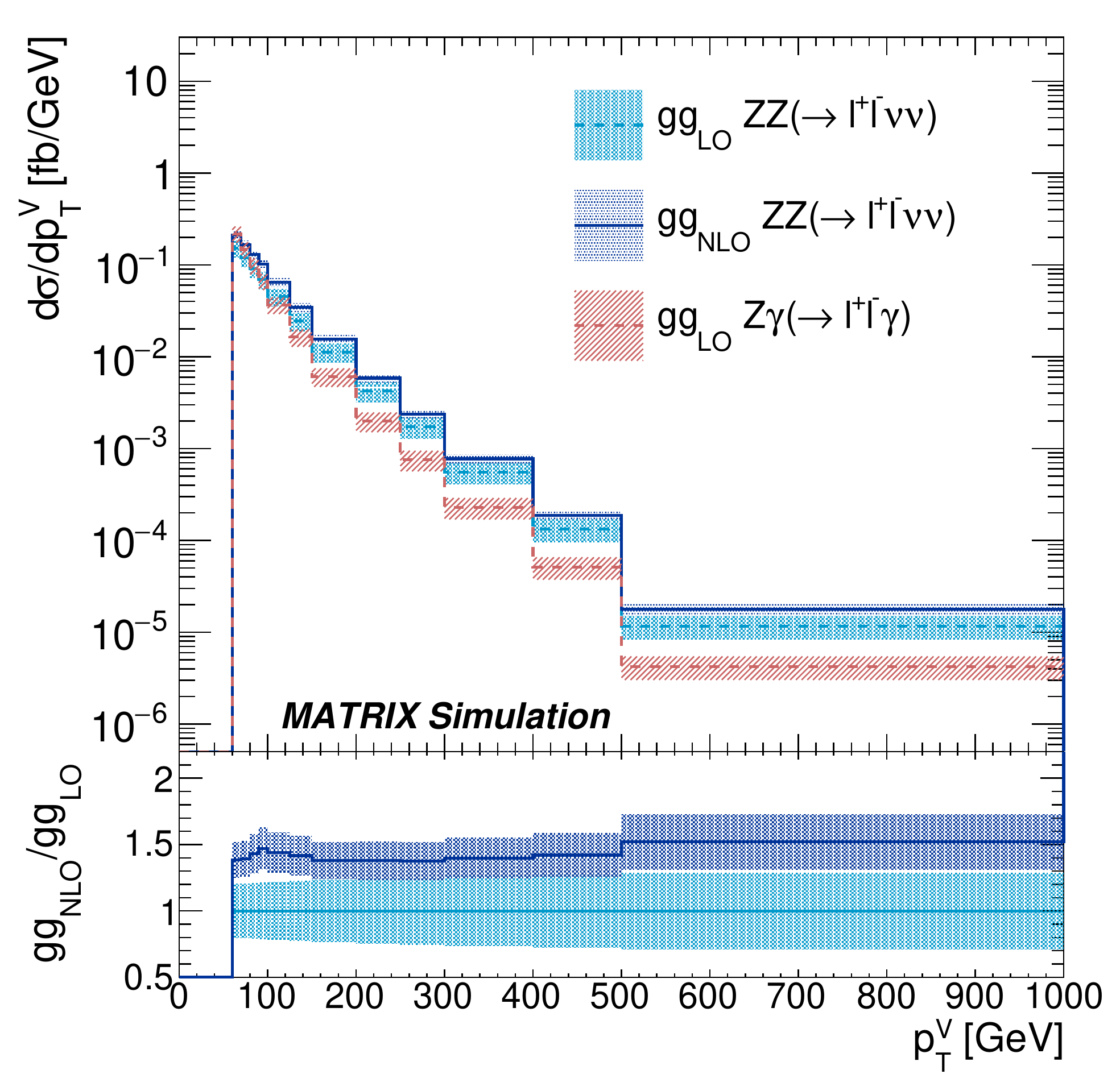}
\caption{}
\end{subfigure}
\hfill
\begin{subfigure}[b]{0.49\textwidth}
\includegraphics[width=\textwidth]{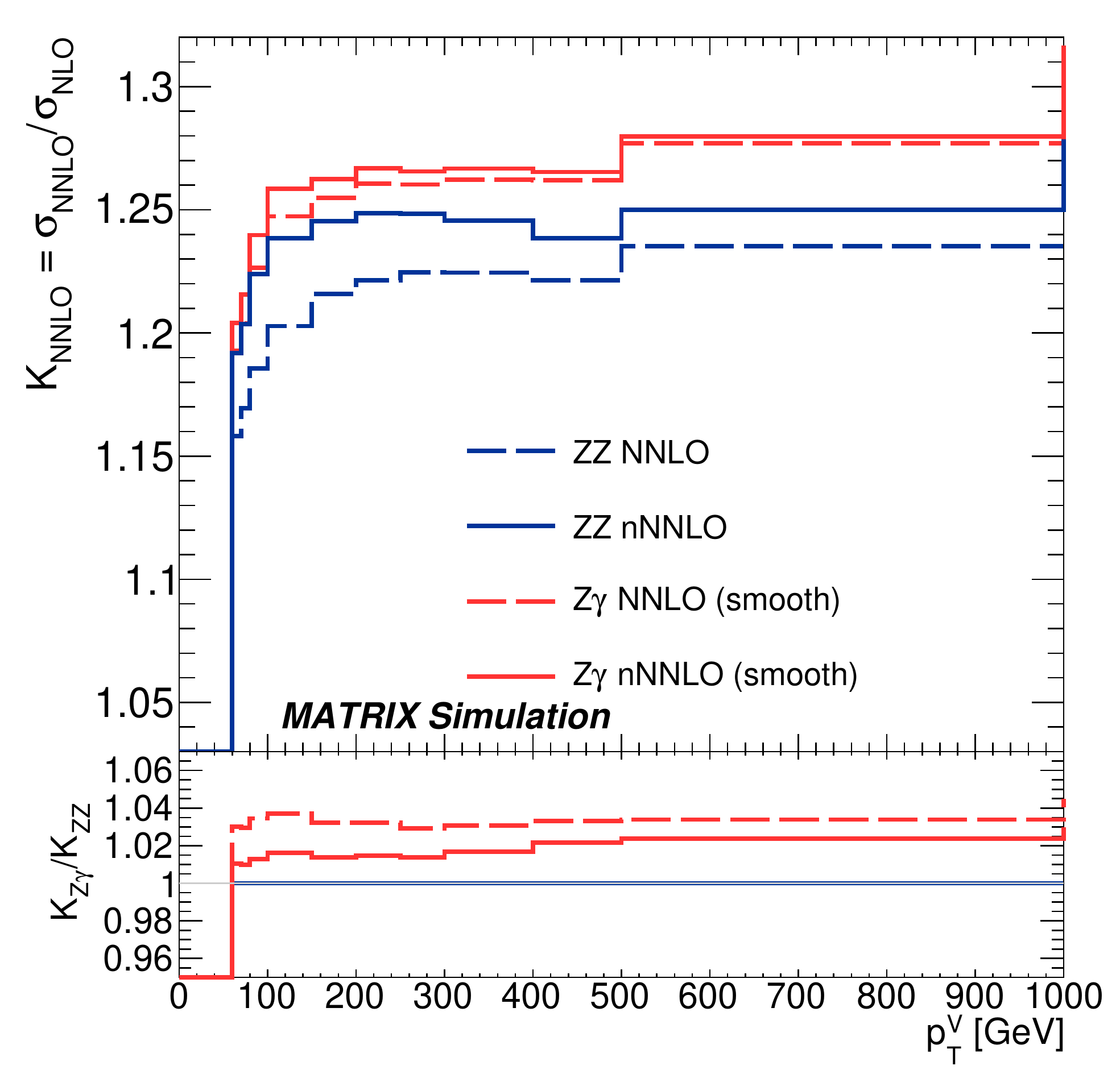}
\caption{}
\end{subfigure}
\caption{(a) \ptV distributions of the $gg \to ZZ(\to \ell^+\ell^-\nu\bar{\nu})$ (blue) and $gg \to Z\gamma(\to \ell^+\ell^-\gamma)$ (red) processes at LO and NLO in QCD together with their scale uncertainty (coloured bands). In the lower panel $K_\mathrm{NLO}$ is shown for $gg \to ZZ(\to \ell^+\ell^-\nu\bar{\nu})$, as well as the scale uncertainty for the LO process. (b) $K$-factor including the $gg$ process at LO (called NNLO) and NLO (called nNNLO) for $ZZ(\to \ell^+\ell^-\nu\bar{\nu})$ and $Z\gamma(\to \ell^+\ell^-\gamma)$ using smooth cone isolation. The lower panel shows the ratios of the $K$-factors for $Z\gamma$ and $ZZ$ production at NNLO (dashed red) and nNNLO (solid red).
}
\label{fig:ggNLO}
\end{figure}

For a fully consistent NNLO calculation of the $pp\to ZZ$ and $pp\to Z\gamma$ processes at ${\cal O}(\alpha_s^2)$, the contributions of the $gg\to ZZ/Z\gamma$ processes need to be included (see Fig.~\ref{fig:ggNLO} (a)). 
The $K$-factors, including both $q\bar{q}$ and $gg$ initial states, are presented in Fig.~\ref{fig:ggNLO} (b). They are shown with the $gg$ process at LO (called NNLO) and at NLO (called nNNLO). The NNLO calculation is consistent at ${\cal O}(\alpha_s^2)$ while the nNNLO includes a subset of the ${\cal O}(\alpha_s^3)$ terms. The relative difference between the $K$-factors for the two processes is considerably reduced compared to the $q\bar{q}$ processes shown in Fig.~\ref{fig:kfactors_nlo_qqnnlo}. The $K$-factor is $3-4$\% higher for $Z\gamma$ production, independent of \ptV. When the NLO $gg$ contributions are included, the difference is further reduced, and is at most 2\% at the highest \ptV. This difference is termed $\delta^\textrm{HO}$ and used as the uncertainty on $R$ due to missing higher-order corrections. 

\begin{figure}[t]
\centering
\includegraphics[width=.69\textwidth]{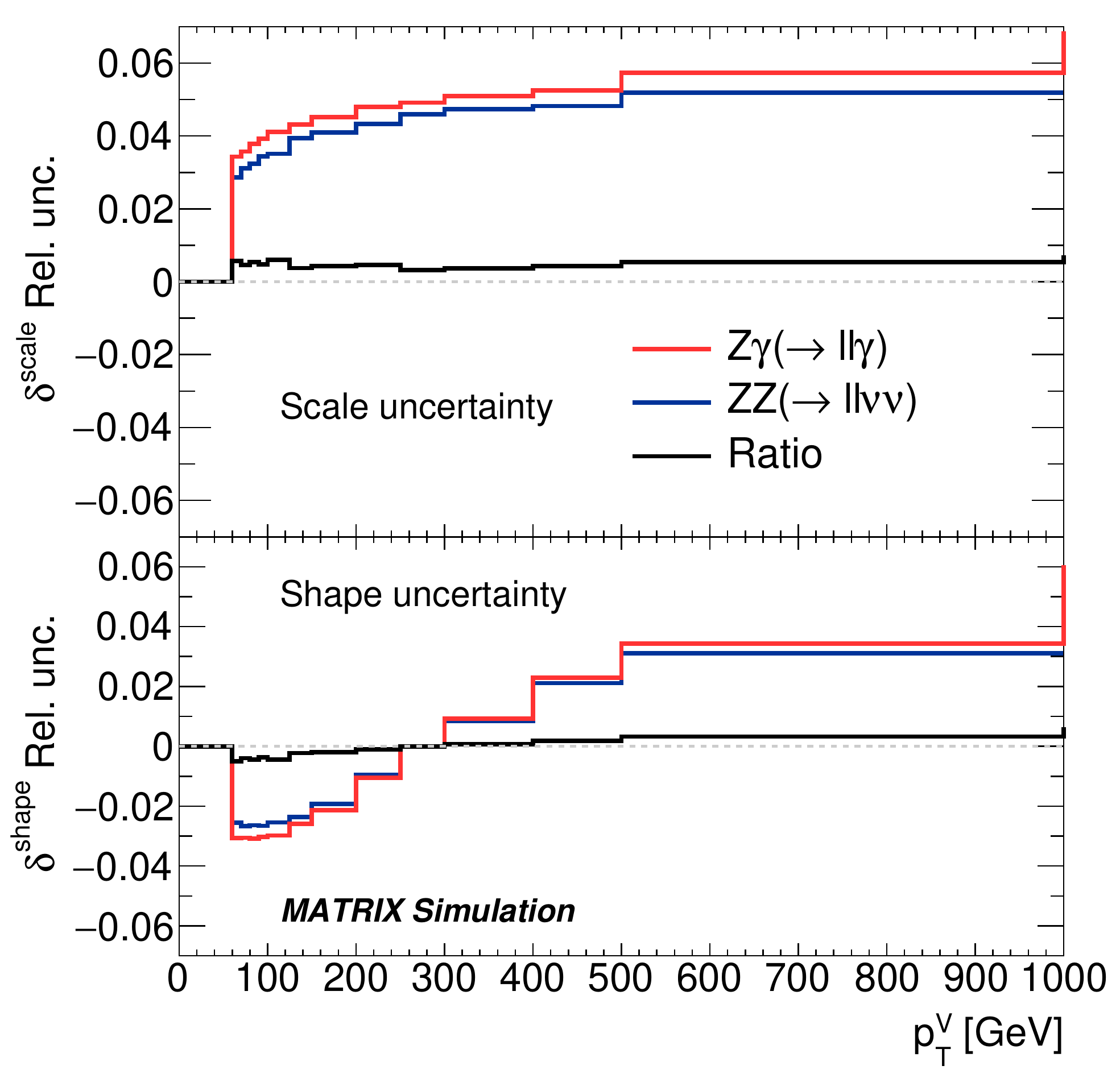} 
\caption{Relative QCD scale (top) and shape (bottom) uncertainties on the $ZZ(\to \ell^+\ell^-\nu\bar{\nu})$ (blue) and $Z\gamma(\to \ell^+\ell^-\gamma)$ (red) cross-sections, and on $R$ (black). 
}

\label{fig:ScaleShape_Unc}
\end{figure}

Two more uncertainties are included, following the procedure outlined in Ref.~\cite{Lindert:2017olm}. The standard scale uncertainty $\delta^\textrm{scale}$ is estimated for the individual cross-sections by varying both the renormalisation and factorisation scales by factors of 2 and 0.5 independently, but discarding variations if the two scales differ by a factor of 4. This leads to seven variations and the cross-section uncertainty is symmetrized by considering half the difference between the lowest and highest value.
To determine the uncertainty on $R$, the same procedure is followed, varying the scales coherently between the two processes. The resulting cross-section uncertainties are about $3-6$\% for the $ZZ$ and $Z\gamma$ processes, while for $R$ they are reduced to about $0.5\%$, as shown in Fig.~\ref{fig:ScaleShape_Unc} (top).
As discussed in Ref.~\cite{Lindert:2017olm},  constant scale variations mainly affect the overall normalisation of \pt distributions and tend to underestimate shape uncertainties, which play an important role in the extrapolation of low-\pt measurements to
high \pt. Thus, for a reasonably conservative estimate of the shape uncertainties, we introduce an additional variation,
$$\delta^\textrm{shape}(\ptV)=\omega^\textrm{shape}(\ptV)\delta^\textrm{scale}(\ptV)$$ 
where $\omega^\textrm{shape}=(\pt^2-p_\textrm{T,0}^2)/(\pt^2+p_\textrm{T,0}^2)$ and $p_\textrm{T,0}=250$~GeV is chosen (roughly in the middle of the relevant part of the \ptV distribution). 
The $\delta^\textrm{shape}$ uncertainty is shown in Fig.~\ref{fig:ScaleShape_Unc} (bottom). It is also much smaller for the ratio $R$ compared to the effect on the individual cross-sections.

\begin{figure}[t]
\centering
\includegraphics[width=.7\textwidth]{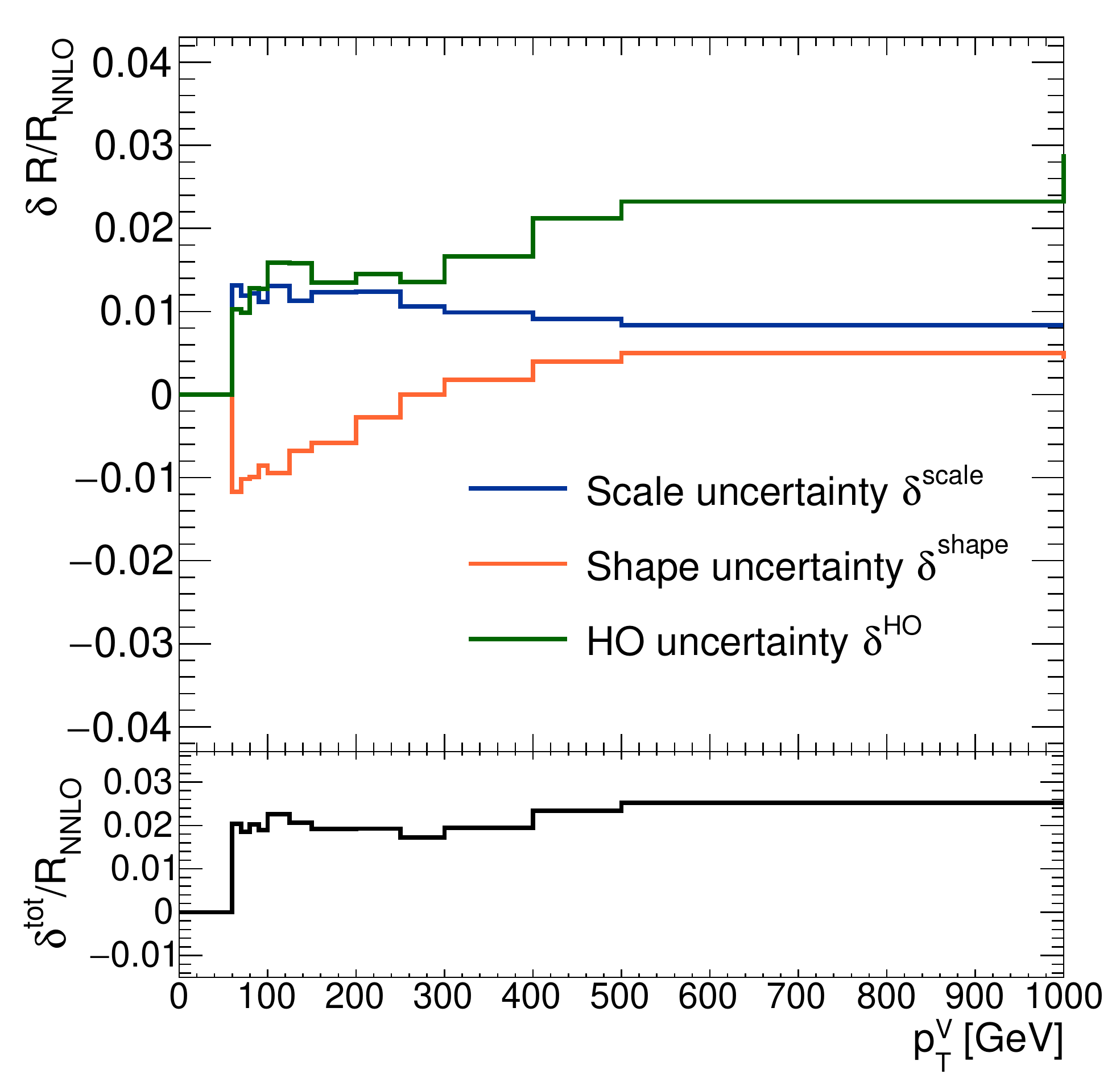}
\caption{Relative uncertainties on $R$ as described in the text: scale (blue) and shape (orange) uncertainties evaluated at NNLO in QCD and the higher-order (HO) uncertainty evaluated at nNNLO (green). 
In the bottom frame the uncertainties are added in quadrature.
}
\label{fig:Relative_QCDUnc}
\end{figure}

The total uncertainty on $R$ due to QCD corrections is the quadratic sum of the three discussed sources:
$$\delta^\textrm{QCD}=\delta^\textrm{HO}\oplus \delta^\textrm{scale}\oplus\delta^\textrm{shape}.$$
It is summarized in Fig.~\ref{fig:Relative_QCDUnc} and amounts to $2-2.5$\%. 

\begin{figure}[t]
\centering
\includegraphics[width=.7\textwidth]{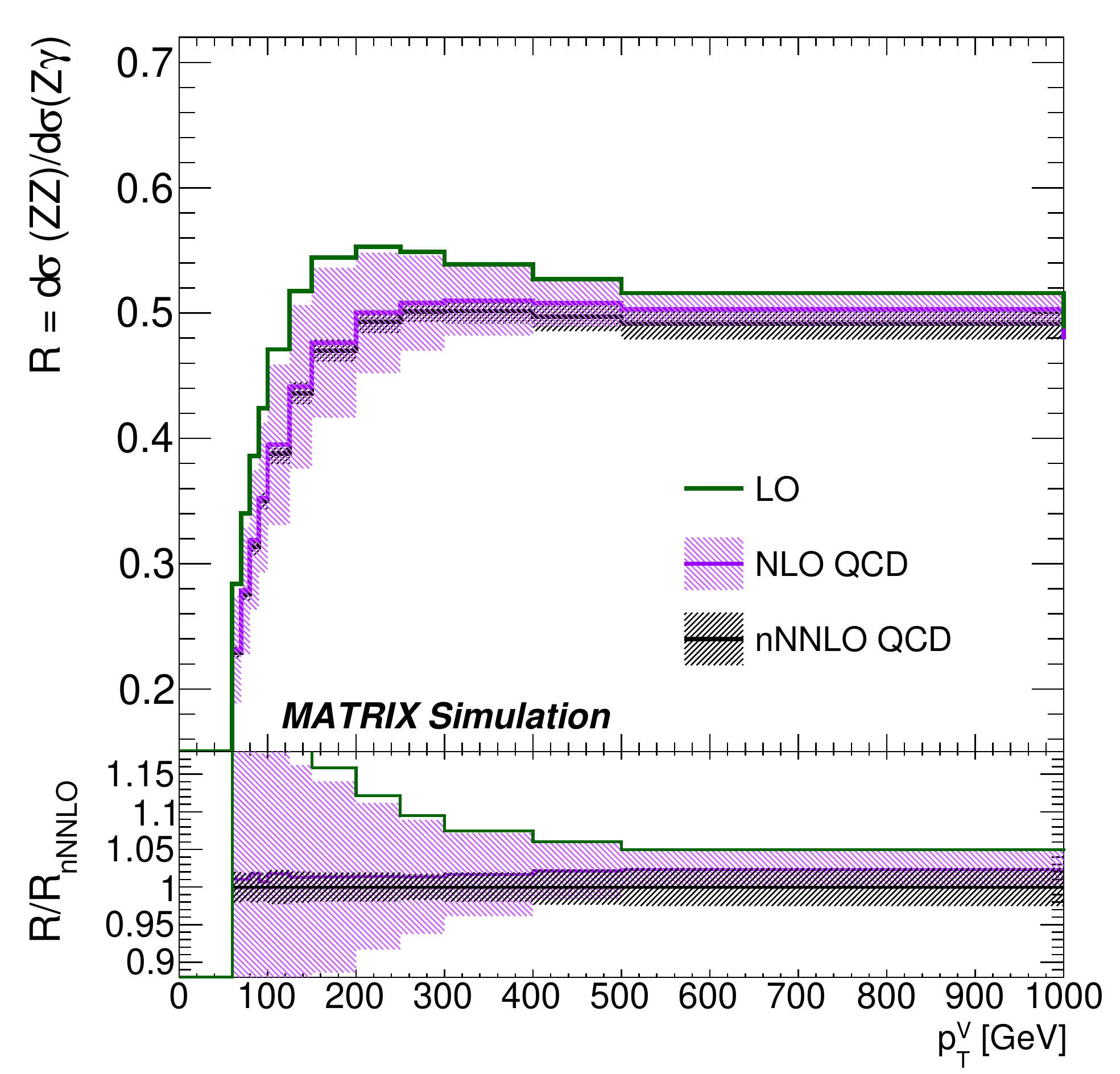}
\caption{Ratio $R$ at LO (green), NLO (purple), nNNLO (gray) in QCD. The bands correspond to the QCD uncertainties ($\delta^{scale},~\delta^{shape},~\delta^{HO}$) added in quadrature. Only $\delta^{scale}$ and $\delta^{shape}$ are considered for the LO case. The bottom frame shows the ratios normalised to the nNNLO prediction.
}
\label{fig:Ratio_QCDUnc}
\end{figure}

Finally, in Fig.~\ref{fig:Ratio_QCDUnc} we show the ratio $R$ at the three different perturbative orders,
 together with the QCD uncertainties.
The $R$-values are very similar at NLO and nNNLO, while the uncertainty is substantially reduced at nNNLO, particularly at low $\ptV$.

\subsection{Isolation}
\label{sec:isolation}
Photon isolation from hadronic activity is required for both theoretical and experimental reasons. While isolation is often used in calculations to avoid divergences, experimentally, the goal is to reject backgrounds from hadron decays produced in fragmentation processes of quarks and gluons.
The smooth cone isolation defined in Eq.~\ref{eq:frixione_iso} is used in cross-section calculations but cannot be applied by the experiments, as hadronic energy around the photon cannot be completely suppressed for a variety of reasons, e.g. to cope with multiple interactions per LHC bunch crossing and electronic noise. Furthermore, the measurement of the hadronic activity around the photon is limited by the detector resolution.
Typically, the ATLAS and CMS collaborations constrain the allowed energy fraction in a cone around the photon with a size that ranges from $R_{0} = 0.2$ to 0.4 depending on the analysis.
Ideally, theoretical and experimental isolations should match, but the required energy fraction requirement can vary between experiments and parts of the detector, so the best choice for the smooth cone parameters is not clear. Previous studies have shown that for a tight enough isolation, the differences between smooth cone and experimental isolation tend to become small \cite{Andersen:2014efa,Catani:2013oma}. Following the recommendations from Ref.~\cite{Andersen:2014efa}, we conclude that the adopted cone parameters are tight enough to proceed.

\begin{figure}[t]
\centering
\begin{subfigure}[b]{0.49\textwidth}
\includegraphics[width=\textwidth]{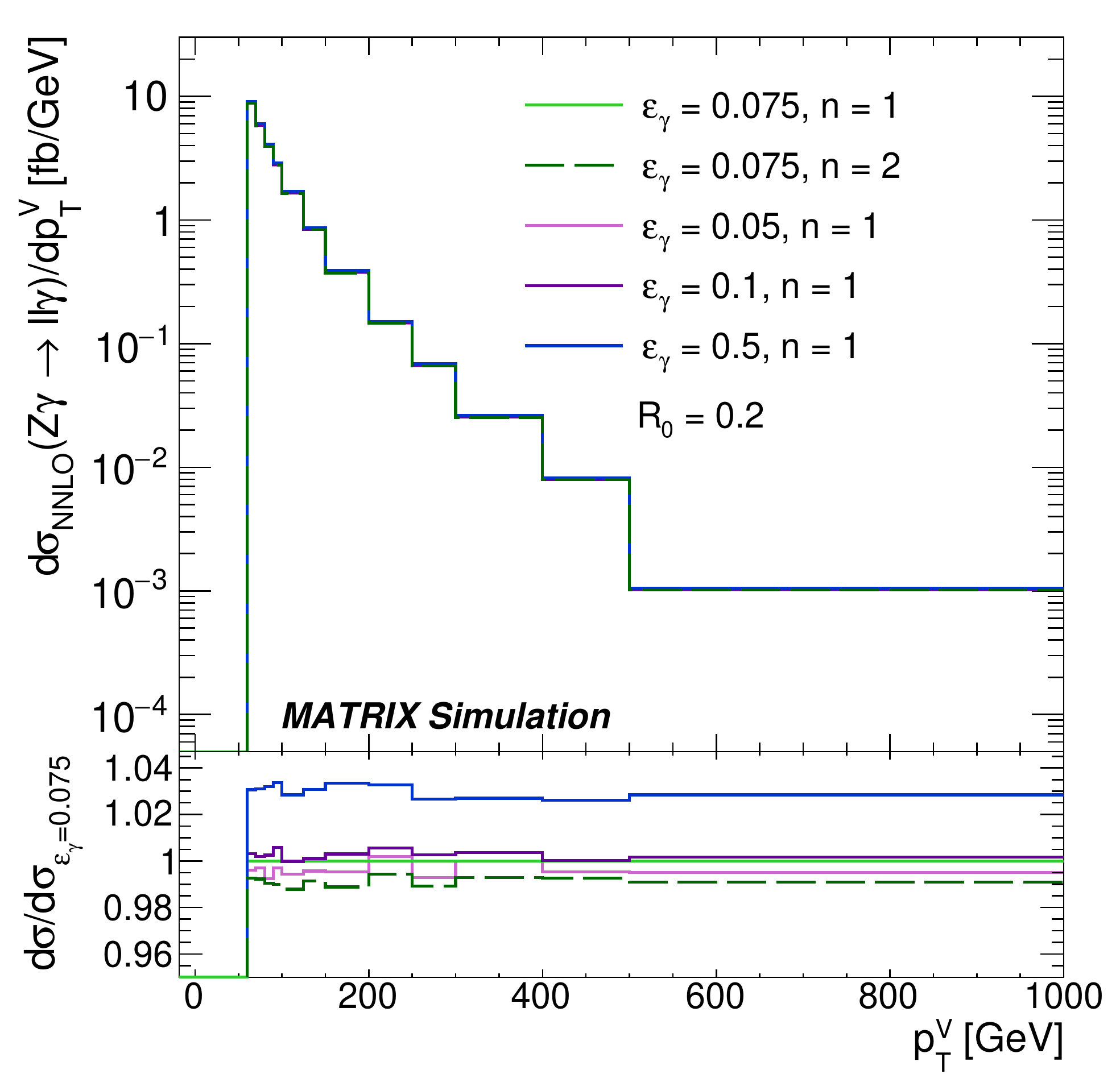} 
\caption{}
\end{subfigure}
\hfill
\begin{subfigure}[b]{0.49\textwidth}
\includegraphics[width=\textwidth]{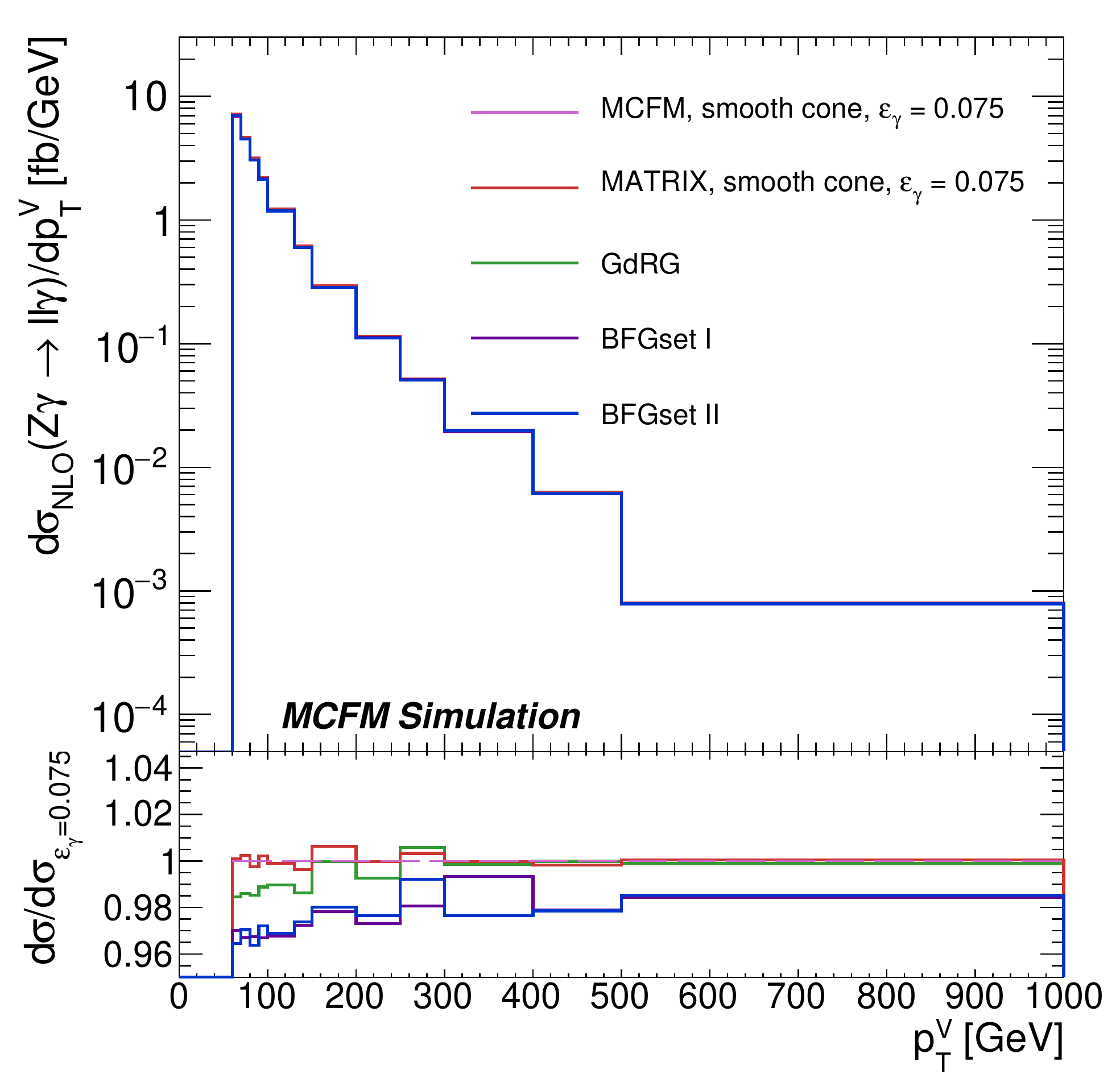}
\caption{}
\end{subfigure}
\caption{Differential $Z\gamma(\to \ell^+\ell^-\gamma)$ cross-sections for different smooth cone isolation parameters (a) and for different fragmentation sets (b). The bottom frames show the cross-sections normalised to the cross-sections obtained with the nominal smooth cone parameters: $\varepsilon_{\gamma} = 0.075, n = 1, R_{0} = 0.2$.}
\label{fig:Isolation_Unc}
\end{figure}

Fig.~\ref{fig:Isolation_Unc} (a) shows how the NNLO cross-sections vary for a fixed $R_{0}=0.2$ and different $\varepsilon_{\gamma}$ and $n$. The curves agree within 0.5\% except for the extreme choice of $\varepsilon_{\gamma}=0.5$ which differs by 3\%. The value $\varepsilon_{\gamma}=0.5$ is indeed very loose and far from the experimental isolation used by ATLAS and CMS. If the parameters are further loosened (decreasing $R_{0},n$ or increasing $\varepsilon_{\gamma}$), the collinear region is encountered giving rise to divergences in the calculation. 

The uncertainty associated with the transition from theoretical to experimental isolation is often estimated by varying the smooth cone parameters~\cite{Campbell:2017dqk,Bern:2011pa}, as done above. We also compare the predictions using different quark-to-photon fragmentation function sets 
based on the \MCFM event generator with NLO precision. 
It was checked that at NLO the \MCFM and \MATRIX results for a given PDF set agree.
Photon fragmentation functions allow to factorize and absorb collinear singularities from final state quarks radiating a highly energetic photon. The fragmentation functions are extracted from fits to experimental data and are only implemented at LO in the \MCFM program\footnote{Previous studies on the diphoton cross-section have shown unphysical results when matching LO fragmentation functions to NLO cross-section calculations~\cite{,Andersen:2014efa,Cieri:2015wwa,Catani:2018krb}.}. Two types of fragmentation functions are considered. They are obtained from the LEP experiments: BFG \cite{Bourhis:1997yu}, GdRG \cite{GehrmannDeRidder:1998ba,GehrmannDeRidder:2006vn}.  

The $Z\gamma$ cross-sections based on different fragmentation functions are compared in Fig.~\ref{fig:Isolation_Unc} (b). 
A discrepancy of $\sim$2\% between the cross-sections based on the GdRG set and the two BFG sets can be seen, as the sets use different phase space regions in the LEP data: BFG corresponds to a more inclusive data set, whereas GdRG is estimated for lower values of the $\epsilon_{\gamma}$ parameter. Comparing the cross-sections based on fragmentation functions and smooth cone isolation, the difference is always lower for the GdRG set, $2\%$ at low \ptV and negligible at high $\ptV$. As discussed in Ref.~\cite{Andersen:2014efa}, a calculation with the smooth cone isolation is more reliable than using fragmentation functions at a lower order. Therefore, a constant uncertainty of $\delta^{iso} = 1\%$ for $Z\gamma$ is applied in the whole $\ptV$ range. 

\subsection{Uncertainty in the Parton Distribution Functions} 
\begin{figure}[h]
\centering
\includegraphics[width=.69\textwidth]{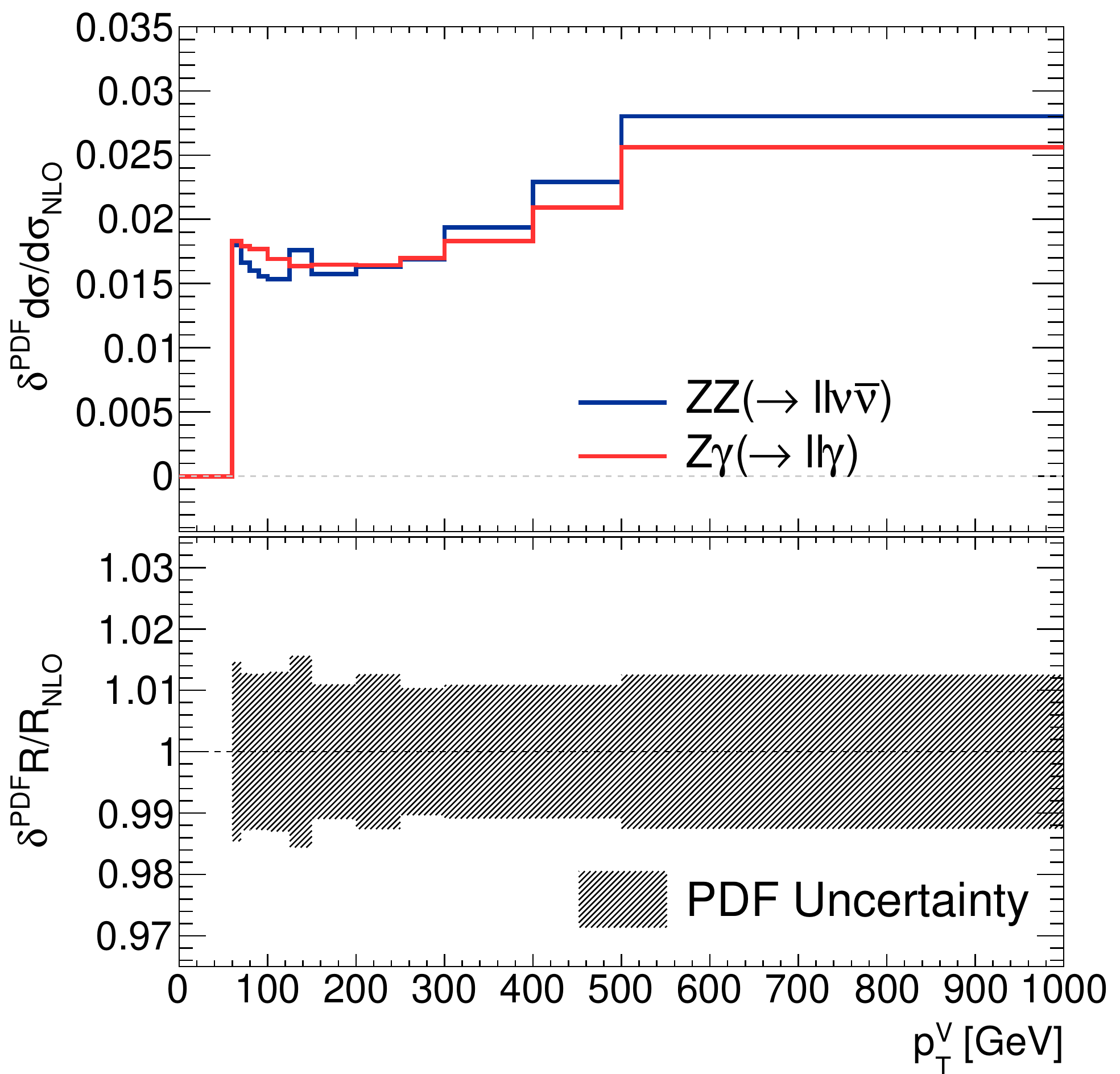}
\caption{Top frame: relative PDF uncertainty on the $ZZ(\to \ell^+\ell^-\nu\bar{\nu})$ (blue) and \hbox{$Z\gamma(\to \ell^+\ell^-\gamma)$} (red) cross-sections. Bottom frame: PDF uncertainty normalised to $R_{NLO}$ (gray band).}
\label{fig:PDF_Unc}
\end{figure}

The uncertainty on $R$ due to the limited knowledge of the PDFs is estimated with the 30 eigenvectors provided by the PDF4LHC15\_30 set~\cite{Butterworth:2015oua}. This uncertainty is evaluated with the \MCFM generator using NLO predictions with NNLO PDFs as follows:
\begin{equation}
\begin{split}
\delta^{PDF} \sigma & = \sqrt{\sum^{N}_{k=1}(\sigma^{(k)} - \sigma^{(0)})^2}, \\
\delta^{PDF} R & = \sqrt{\sum^{N}_{k=1}(R^{(k)} - R^{(0)})^2},
\end{split}
\label{eq:PDFeq}
\end{equation}
where $N$ corresponds to the number of PDF sets, in our case $N = 30$. $\sigma^{(k)},R^{(k)}$ and $\sigma^{(0)},R^{(0)}$ are the cross-sections and ratios evaluated for each set and for the nominal PDF set, respectively. Fig.~\ref{fig:PDF_Unc} shows the PDF uncertainty for both the cross-sections and the ratio, resulting in $\sim$$2-3\%$ and $\sim$$1\%$ in the whole \ptV range, respectively.

\section{Effects of additional selection criteria}
\label{sec:cuts}

Searches at the LHC usually include selection criteria based on event topologies in order to better suppress events from background processes. These can affect $ZZ$ and $Z\gamma$ events differently and should be applied when calculating the cross-section ratio and evaluating the QCD uncertainties. 
 Table~\ref{tab:FullSRdef} shows the additional criteria chosen for this article, which approximate the selection in ATLAS $\ell\ell + \MET$ searches~\cite{Aaboud:2017rel,Aaboud:2017bja}. 
They include angular selections on the leptons and $\MET$, as well as a requirement on the ``missing $E_T$ significance'', $\MET / \sqrt{\sum E_T}$, which for $Z\gamma$ corresponds to $\ptgamma / \sqrt{\sum E_T}$. Here, $\sum E_T$ is the scalar sum of the transverse momenta of the leptons and hadronic jets in the event, where the jets pass $\pt > 20$~\GeV\ and $|\eta| < 4.5$. Due to the existing $\MET$ and $\ptgamma$ selections, this requirement effectively vetoes events with significant hadronic activity.

\begin{table}[h!]
\begin{center}
\begin{tabular}{l | c | c}
\noalign{\smallskip}\hline\noalign{\smallskip}
   Variable & $ZZ$ & $Z\gamma$ \\
 \noalign{\smallskip}\hline\noalign{\smallskip}
 $\Delta R (\ell_1,\ell_2)$ &  \multicolumn{2}{c}{$<1.8$} \\
 $\Delta \phi (Z,\MET)$ & $>2.7$ & -  \\
 $\Delta \phi (Z,\gamma)$ & - & $>2.7$  \\
 $\MET / \sqrt{\sum E_T}$ & $>9$ & - \\
 $\ptgamma / \sqrt{\sum E_T}$ & - & $>9$ \\
\hline
\end{tabular}
\end{center}
\caption{\label{tab:FullSRdef} Additional requirements on \ZZllnunu (left column) and $\Zg$ (right column) events, similar to those used in ATLAS $\ell\ell + \MET$ searches for physics beyond the SM.}
\end{table}

\begin{figure}[h]
\centering
\begin{subfigure}[b]{0.49\textwidth}
\includegraphics[width=\textwidth]{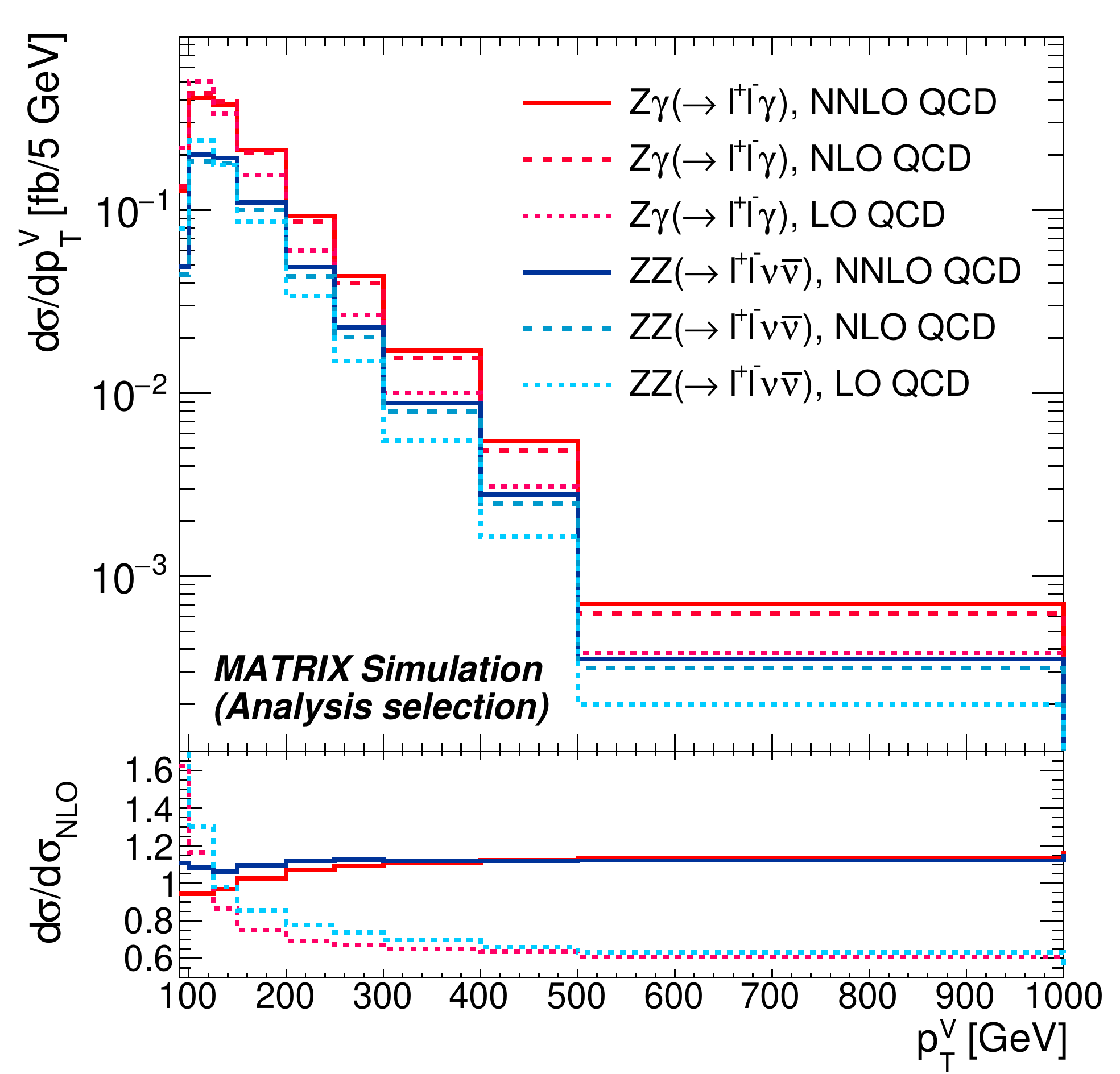}
\caption{}
         \label{fig:xsa_cuts}
\end{subfigure}
\hfill
\begin{subfigure}[b]{0.49\textwidth}
\includegraphics[width=\textwidth]{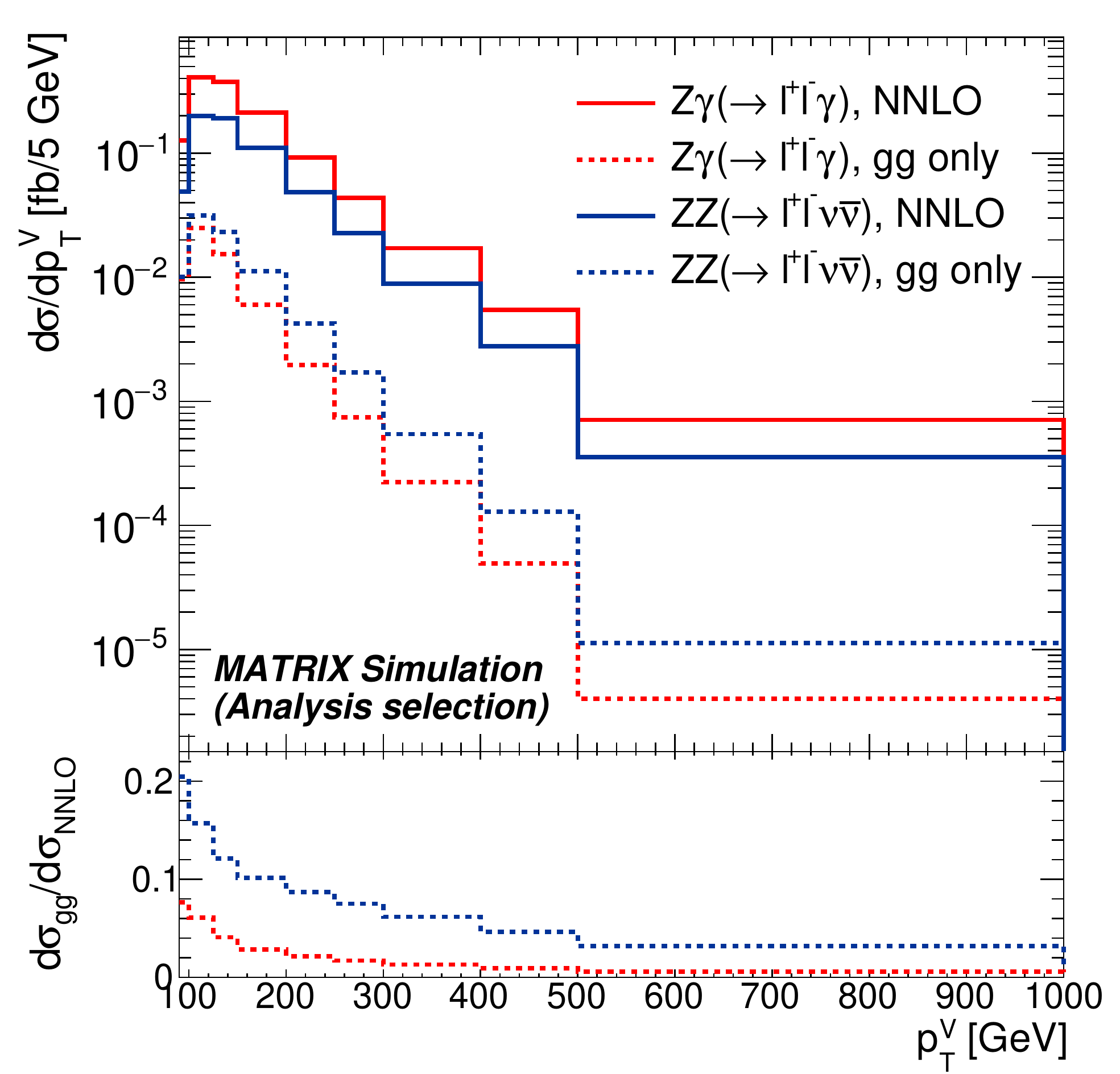}
\caption{}
         \label{fig:xsb_cuts}
\end{subfigure}
\caption{(a) $ZZ (\to \ell^+\ell^-\nu\bar{\nu})$ and $Z\gamma(\to \ell^+\ell^-\gamma)$ cross-sections as a function of \ptV at LO (dotted), NLO (dashed) and NNLO (solid) in QCD. The bottom frame shows the ratio of the LO and NNLO calculations to the NLO prediction. (b) \ptV distribution at NNLO in QCD (solid) and for the $gg$-induced contribution separately (dotted). The bottom frame shows the fractional contribution of the $gg$ process. The additional selection detailed in Table~\ref{tab:FullSRdef} is applied for both distributions.
}
\label{fig:zzzgxs_cuts}
\end{figure}

Figure~\ref{fig:zzzgxs_cuts} shows the $ZZ$ and $Z\gamma$ cross-sections as a function of \ptV at different orders in QCD, as well as the $gg$ fraction. In comparison to the distributions in Fig.~\ref{fig:zzzgxs}, the additional selection reduces the integrated cross-section by about a factor of 5 for $ZZ$ and a factor of 7 for $Z\gamma$. Additionally, the relative fraction of $gg$-initiated events is increased by roughly a factor of 1.5 for both the $ZZ$ and $Z\gamma$ processes. In these and the following figures, only results for \ptV $>$ 90~\GeV\ are shown, as there are very few events below this value after the additional selection is applied.

\begin{figure}[h]
\centering
\includegraphics[width=.7\textwidth]{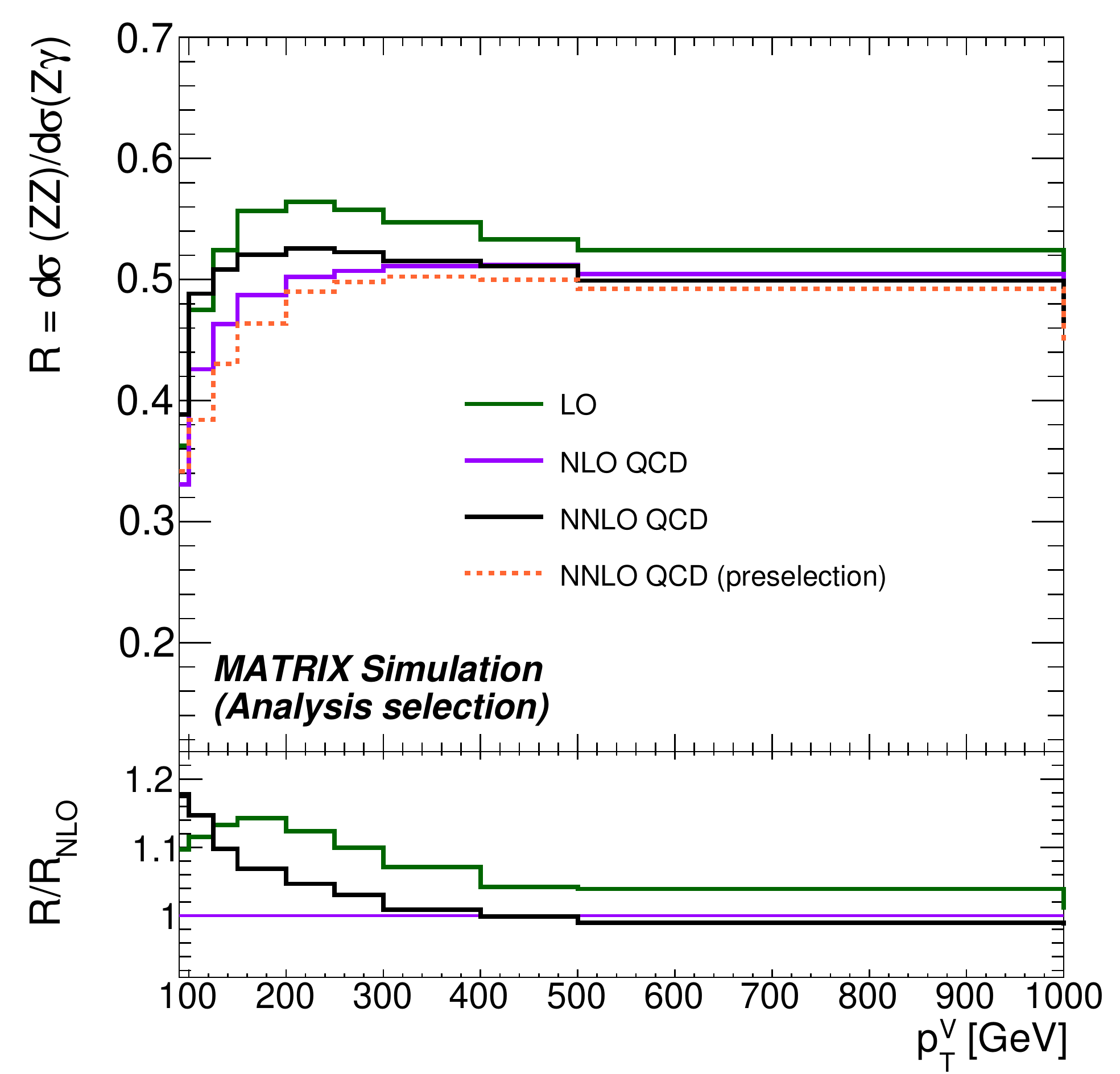} 
\caption{Ratio $R$ between the $ZZ(\to \ell^+\ell^-\nu\bar{\nu})$ and $Z\gamma(\to \ell^+\ell^-\gamma)$ differential cross-sections as function of \ptV at LO (green), NLO (violet) NNLO (black) in QCD, with the additional selection detailed in Table~\ref{tab:FullSRdef}, and the comparison to the NNLO ratio after preselection only (orange dotted). The bottom plot shows the ratios relative to the NLO calculation.}
\label{fig:Ratio_cuts}
\end{figure}

The cross-section ratio $R$ is shown in Fig.~\ref{fig:Ratio_cuts}. The additional selection increases the NNLO ratio compared to the preselection result in the entire considered \ptV range, and particularly for $\ptV < $ 300~\GeV. In contrast to the preselection result (see Fig.~\ref{fig:Ratio}), $R$ calculated at NLO is lower than $R$ at NNLO for $\ptV < 400$~\GeV. 

\subsection{Effects of additional selection criteria on QCD uncertainties}
\label{subsec:qcdunc_cuts}

The same methodologies are applied for the evaluation of the systematic uncertainties on $R$ as in Section~\ref{sec:qcdunc}. For the QCD uncertainties, these are higher-order uncertainties, as well as scale variations and shape uncertainties. The PDF and isolation uncertainties are not expected to change significantly compared to those evaluated for the preselection. %

\begin{figure}[t]
\centering
\begin{subfigure}[b]{0.49\textwidth}
\includegraphics[width=\textwidth]{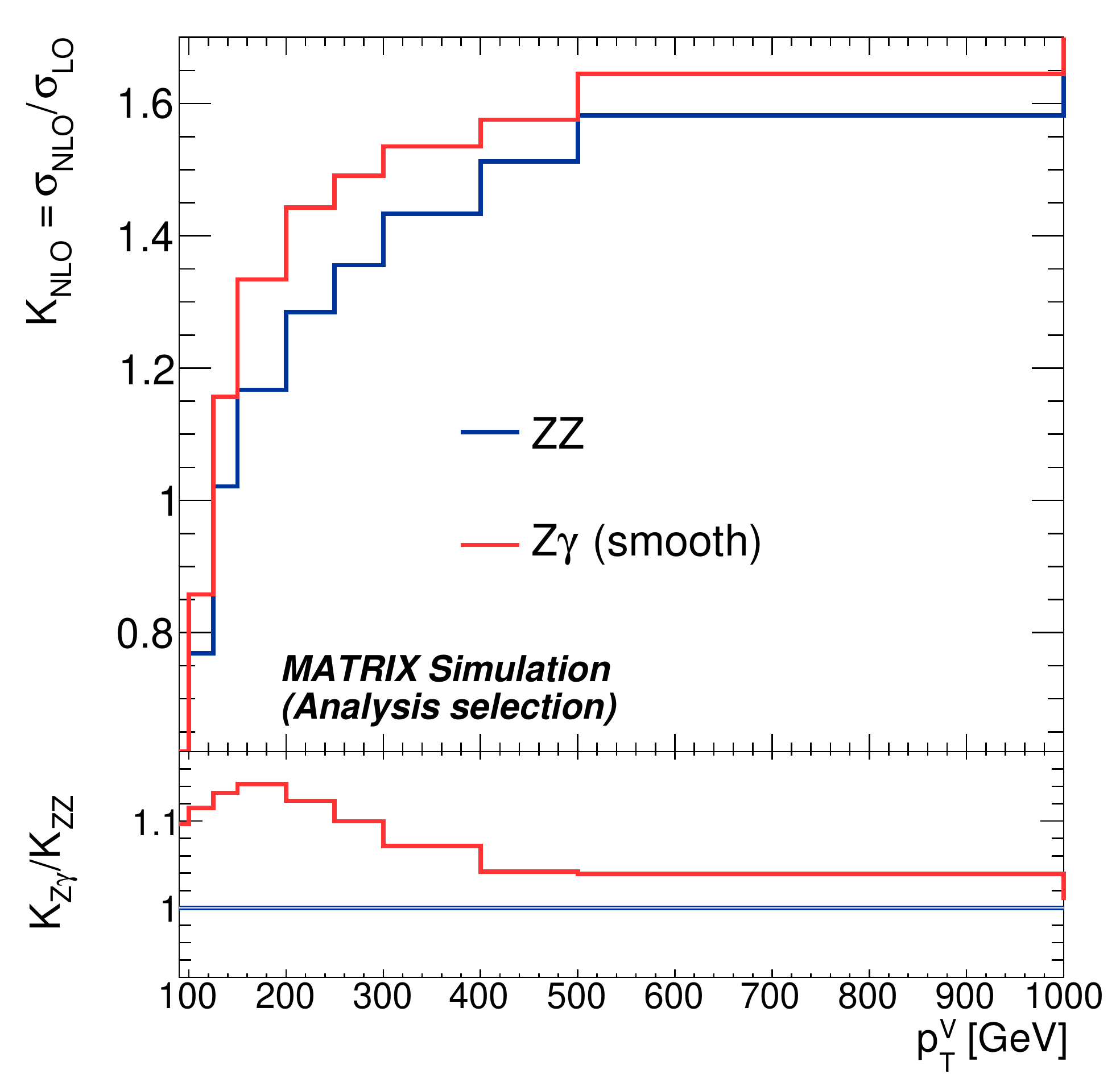}
\caption{}
\end{subfigure}
\hfill
\begin{subfigure}[b]{0.49\textwidth}
\includegraphics[width=\textwidth]{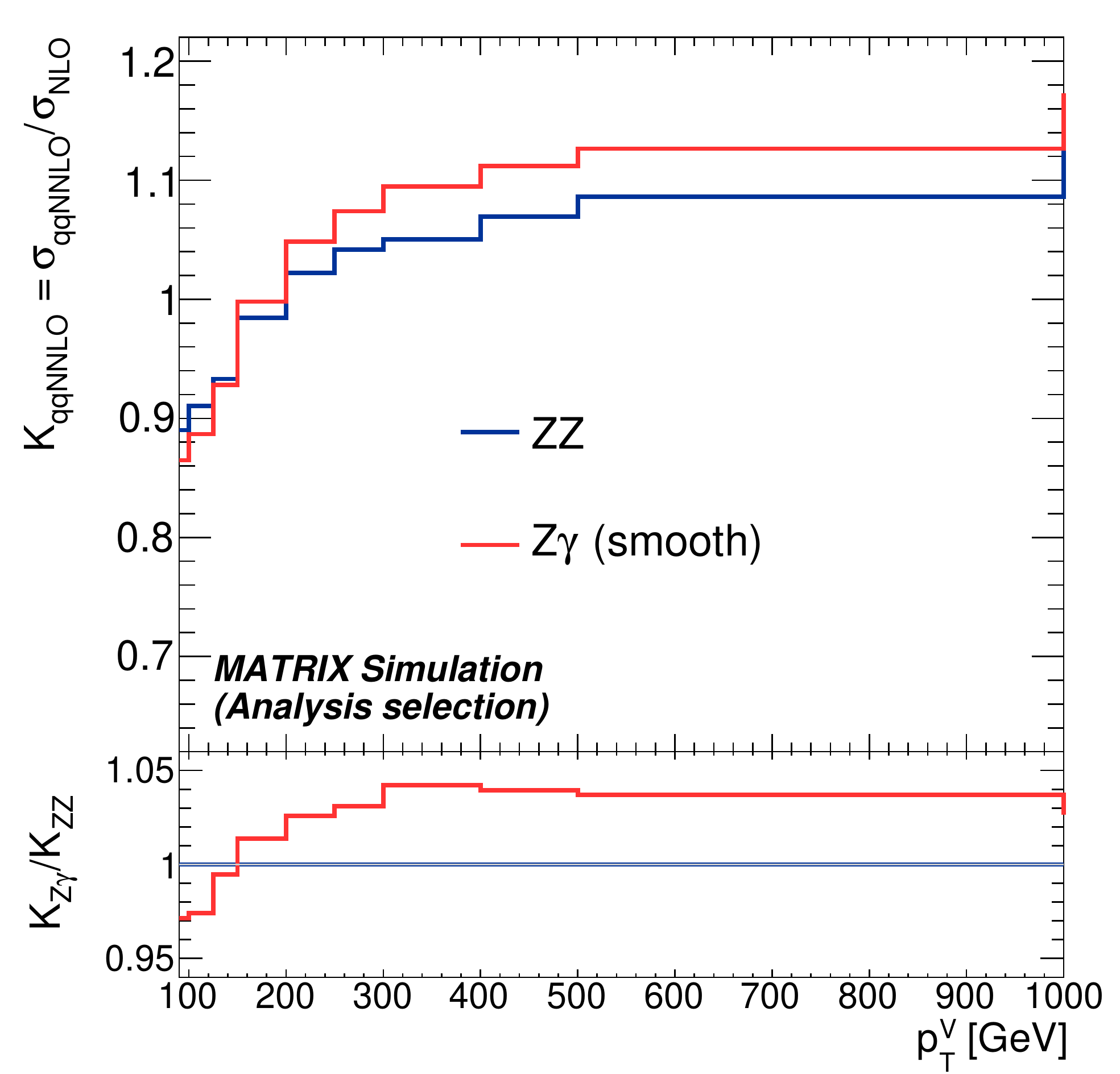} 
\caption{}
\end{subfigure}
\caption{$K_{\rm NLO}$ (a) and $K_{\rm NNLO}$ (b) for $q\bar{q}\to ZZ(\to \ell^+\ell^-\nu\bar{\nu})$ (blue) and \hbox{$q\bar{q}\to Z\gamma(\to \ell^+\ell^-\gamma)$} with smooth cone isolation (red). The bottom panels show the ratio between the $K$-factors for $Z\gamma$ and $ZZ$ production. The selection detailed in Table~\ref{tab:FullSRdef} is applied in all distributions.
}
\label{fig:kfactors_nlo_qqnnlo_cuts}
\end{figure}

Figure~\ref{fig:kfactors_nlo_qqnnlo_cuts} shows the NLO and NNLO $K$-factors of the $q\bar{q} \rightarrow ZZ$ and $q\bar{q} \rightarrow Z\gamma$ processes. Compared to Fig.~\ref{fig:kfactors_nlo_qqnnlo}, the additional selection decreases the $K$-factors for both processes, with a larger effect on $Z\gamma$. The $K$-factors are now more similar between the two processes than in the case of the preselection. 

\begin{figure}[t]
\centering
\begin{subfigure}[b]{0.49\textwidth}
\includegraphics[width=\textwidth]{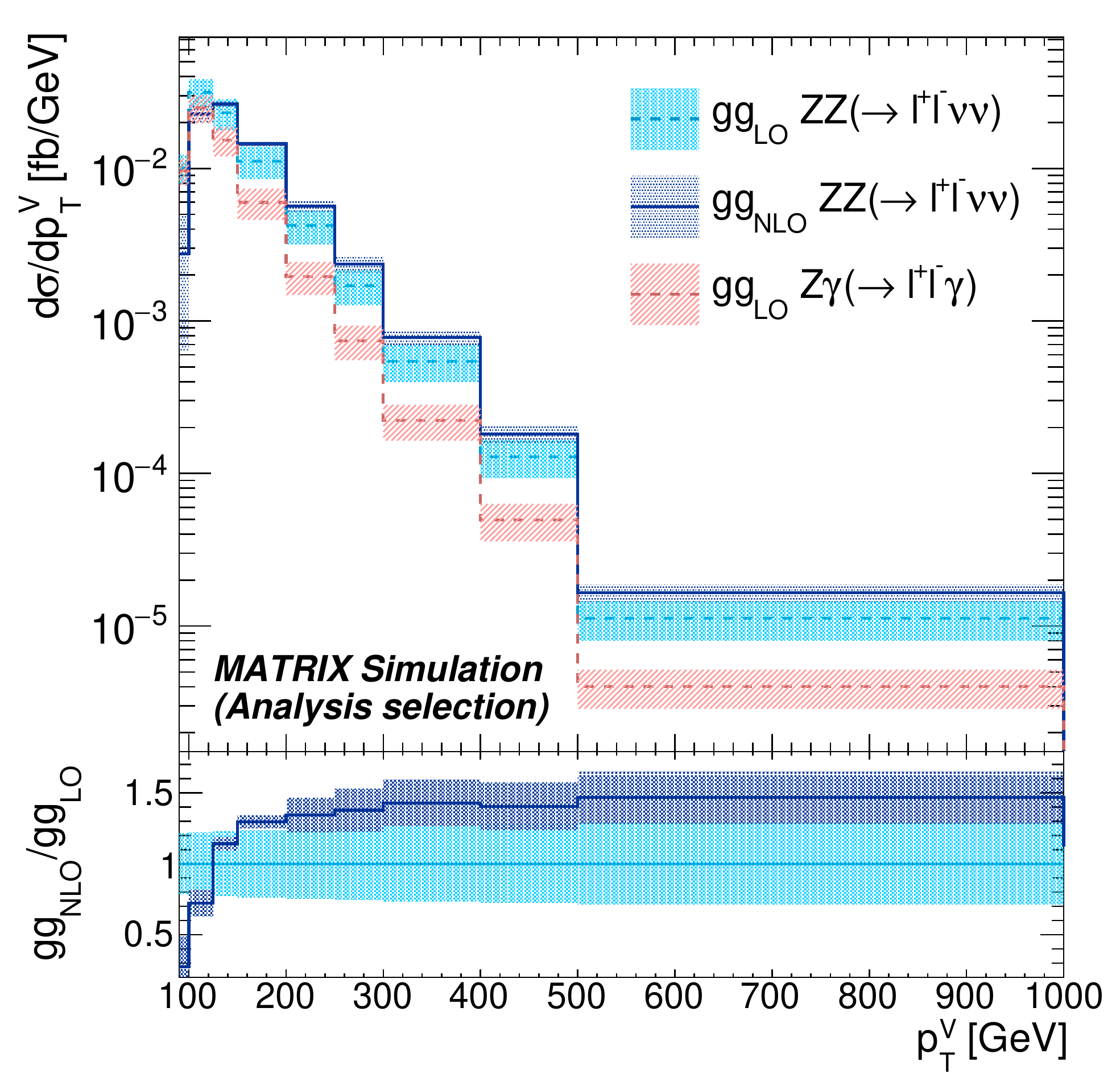}
\caption{}
\end{subfigure}
\hfill
\begin{subfigure}[b]{0.49\textwidth}
\includegraphics[width=\textwidth]{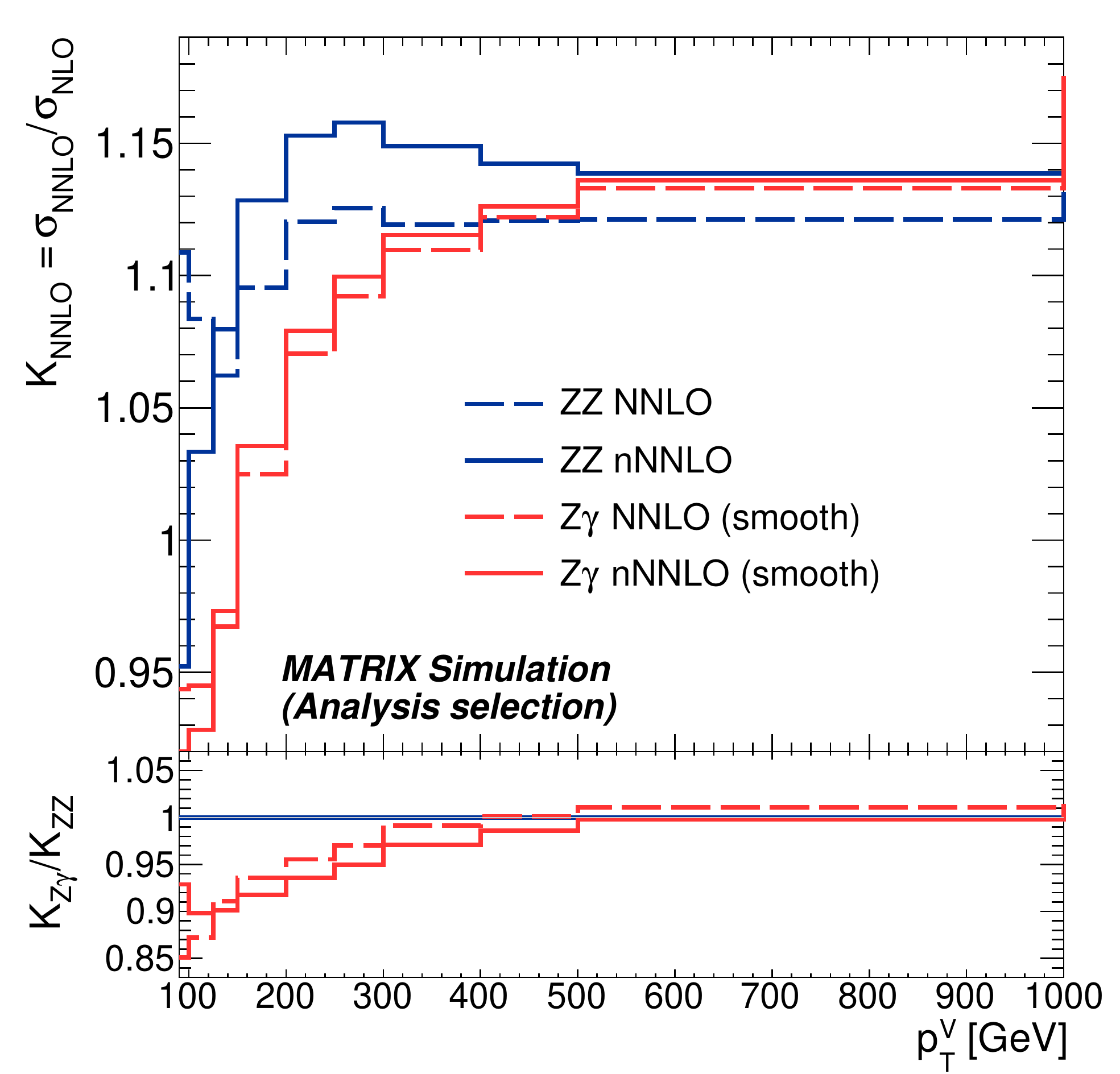}
\caption{}
\end{subfigure}
\caption{(a) \ptV distributions of the $gg \to ZZ(\to \ell^+\ell^-\nu\bar{\nu})$ (blue) and {$gg \to Z\gamma(\to \ell^+\ell^-\gamma)$} (red) processes at LO and NLO in QCD together with their scale uncertainty (coloured bands).  In the lower panel $K_\mathrm{NLO}$ is shown for $gg \to ZZ(\to \ell^+\ell^-\nu\bar{\nu})$, as well as the scale uncertainty for the LO process.  (b) $K$-factor including the $gg$ process at LO (called NNLO) and NLO (called nNNLO) for $ZZ(\to \ell^+\ell^-\nu\bar{\nu})$ and $Z\gamma(\to \ell^+\ell^-\gamma)$ using smooth cone isolation. The lower panel shows the ratios of the $K$-factors for $Z\gamma$ and $ZZ$ production at NNLO (dashed red) and nNNLO (solid red). The selection detailed in Table~\ref{tab:FullSRdef} is applied in all distributions.}
\label{fig:ggNLO_cuts}
\end{figure}

Again, for the full NNLO calculation at ${\cal O}(\alpha_s^2)$ the $gg$-initiated process needs to be considered, as shown in Fig.~\ref{fig:ggNLO_cuts}. The $K$-factor based on the full NNLO calculation is larger for $ZZ$ than for $Z\gamma$ production. 
In particular, at low $\ptV$, the $K$-factor is about 10\% higher. As discussed earlier, the NLO corrections to the \hbox{$gg\to ZZ$} process are applied to both the \hbox{$gg\to ZZ$} and the \hbox{$gg\to Z\gamma$} process to obtain an nNNLO prediction, which is also shown in Fig.~\ref{fig:ggNLO_cuts}. Above $\ptV > 130$~\GeV, this correction increases both $K$-factors by less than 3\%, and does not significantly change their ratio. Thus, the higher-order uncertainties, derived by evaluating the similarity between the $ZZ$ and $Z\gamma$ $K$-factors at nNNLO, are much higher  at low $\ptV$ for the additional selection compared to the baseline selection. Only for $\ptV>300$~GeV, they reach $\sim$3$\%$, similar to the uncertainties after the baseline selection.
One reason is the relative contribution of the $gg$ process, which is bigger for $ZZ$ production, and constitutes a larger fraction of the total cross-section when the additional selection is applied (see Fig.~\ref{fig:zzzgxs_cuts}~(b)). Furthermore, the additional selection suppresses hadronic activity, therefore causing an imbalance of real and virtual corrections.
Due to the difference in the boson mass and the different contribution of the gluon-initiated production, the $ZZ$ and $Z\gamma$ processes are affected differently.

\begin{figure}[h]
\centering
\begin{subfigure}[b]{0.49\textwidth}
\includegraphics[width=\textwidth]{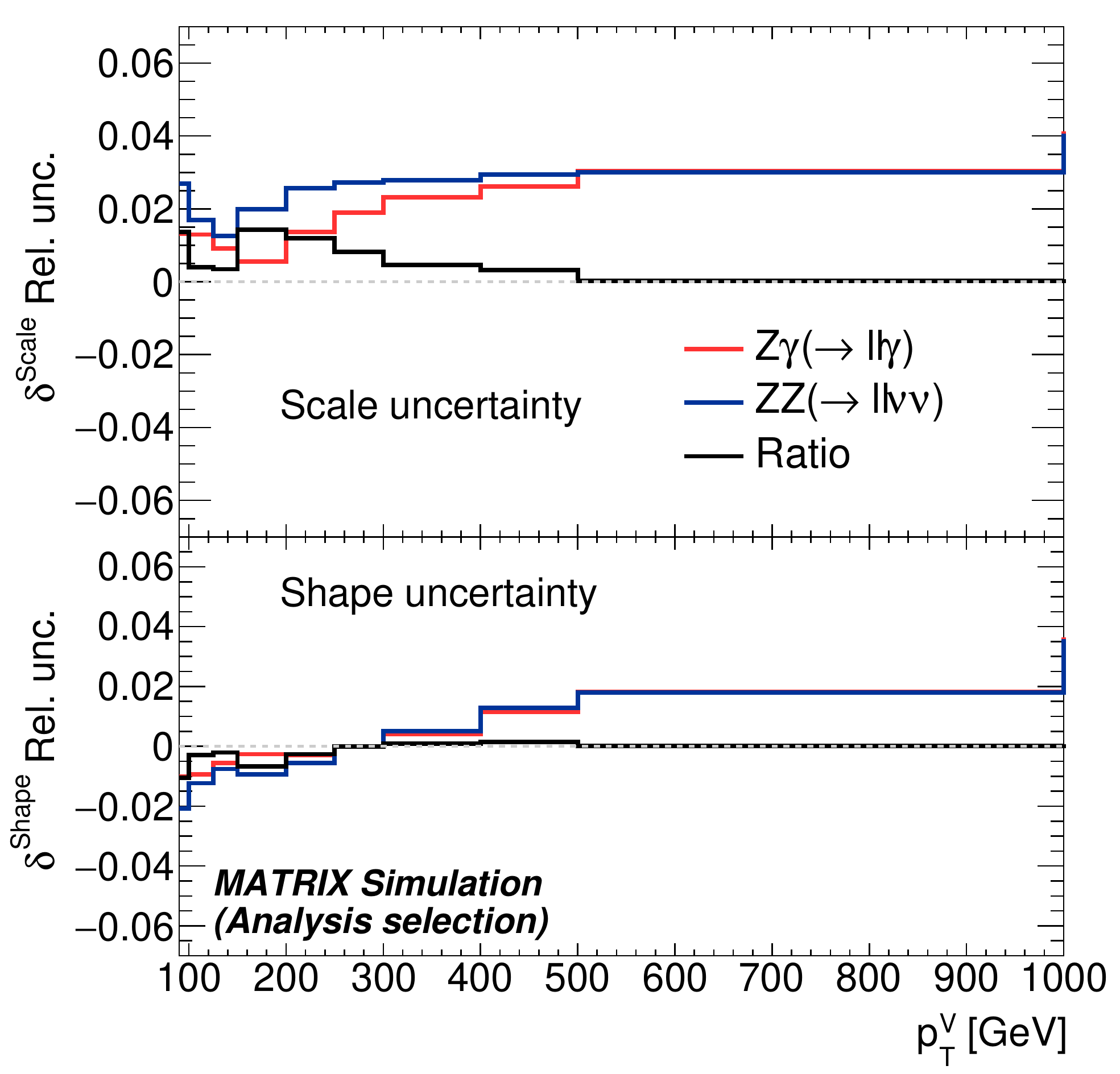} 
\caption{}
\label{fig:ScaleShape_Unc_cuts}
\end{subfigure}
\hfill
\begin{subfigure}[b]{0.49\textwidth}
\includegraphics[width=\textwidth]{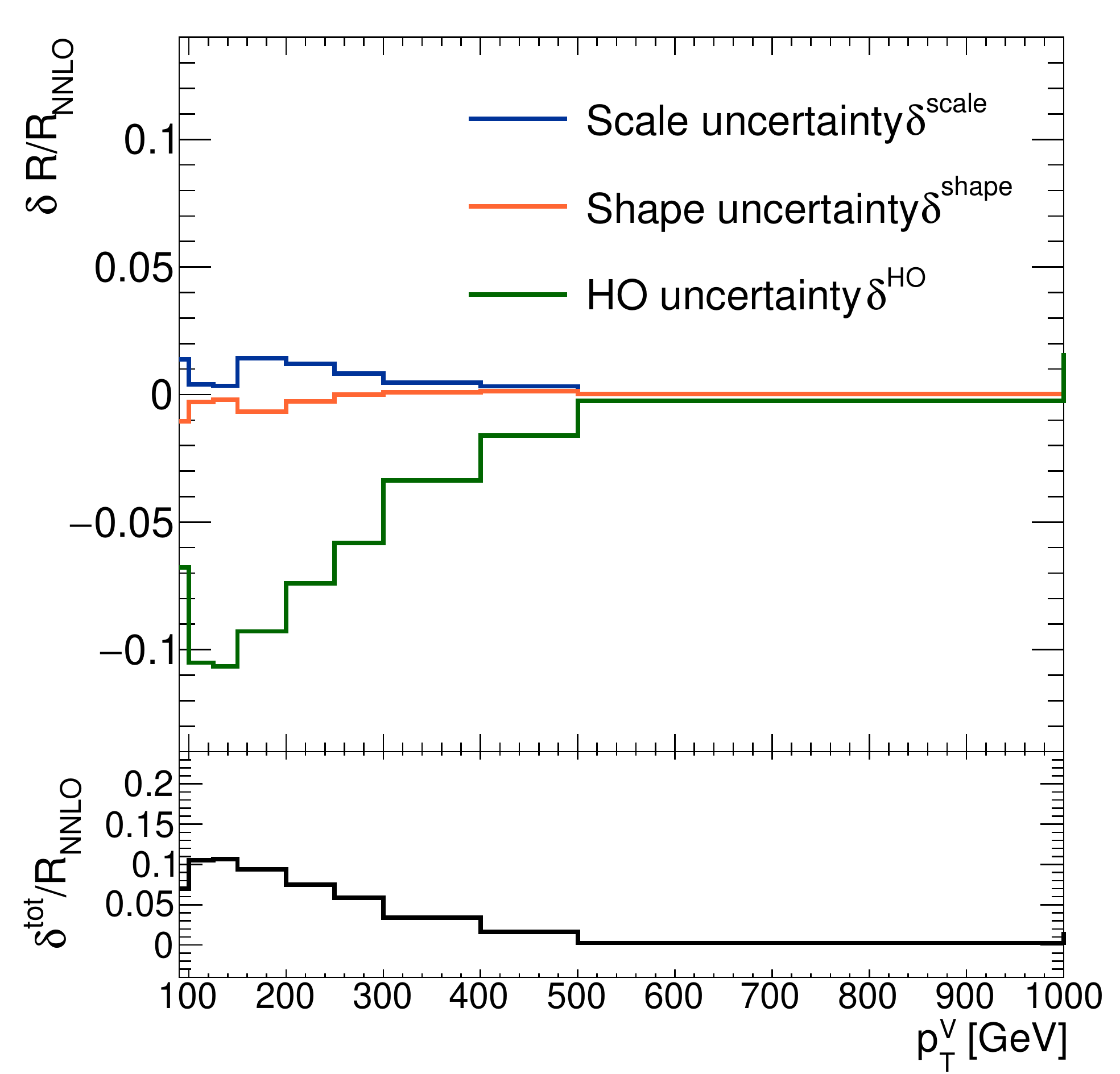}
\caption{}
\label{fig:Relative_QCDUnc_cuts}
\end{subfigure}
\caption{(a) Relative QCD scale (top) and shape (bottom) uncertainties on the $ZZ(\to \ell^+\ell^-\nu\bar{\nu})$ (blue) and $Z\gamma(\to \ell^+\ell^-\gamma)$ (red) cross-sections, and on $R$ (black). (b) Relative uncertainties on $R$ as described in the text: QCD scale (blue) and shape (orange) uncertainties evaluated at NNLO, and higher-order (HO, green) uncertainties evaluated at nNNLO. In the bottom frame the uncertainties are added in quadrature. 
The selection detailed in Table~\ref{tab:FullSRdef} is applied in all distributions.
}
\label{fig:ScaleShape_Unc_and_QCD_cuts}
\end{figure}

The scale and shape uncertainties are presented in Fig.~\ref{fig:ScaleShape_Unc_and_QCD_cuts}~(a). Their size is comparable to the uncertainties at preselection level (Fig.~\ref{fig:ScaleShape_Unc}). 

Figure~\ref{fig:ScaleShape_Unc_and_QCD_cuts} (b) shows a summary of the three QCD uncertainties. They are dominated by the higher-order uncertainty, determined from the nNNLO $K$-factor difference between the $ZZ$ and $Z\gamma$ processes. 
It is possible that a smaller uncertainty can be achieved by performing a multi-dimensional analysis. 

\FloatBarrier

\section{Electroweak corrections and summary of uncertainties}
\label{sec:ewkcor}

\begin{figure}[h]
\centering
\begin{subfigure}[b]{0.49\textwidth}
\includegraphics[width=\textwidth]{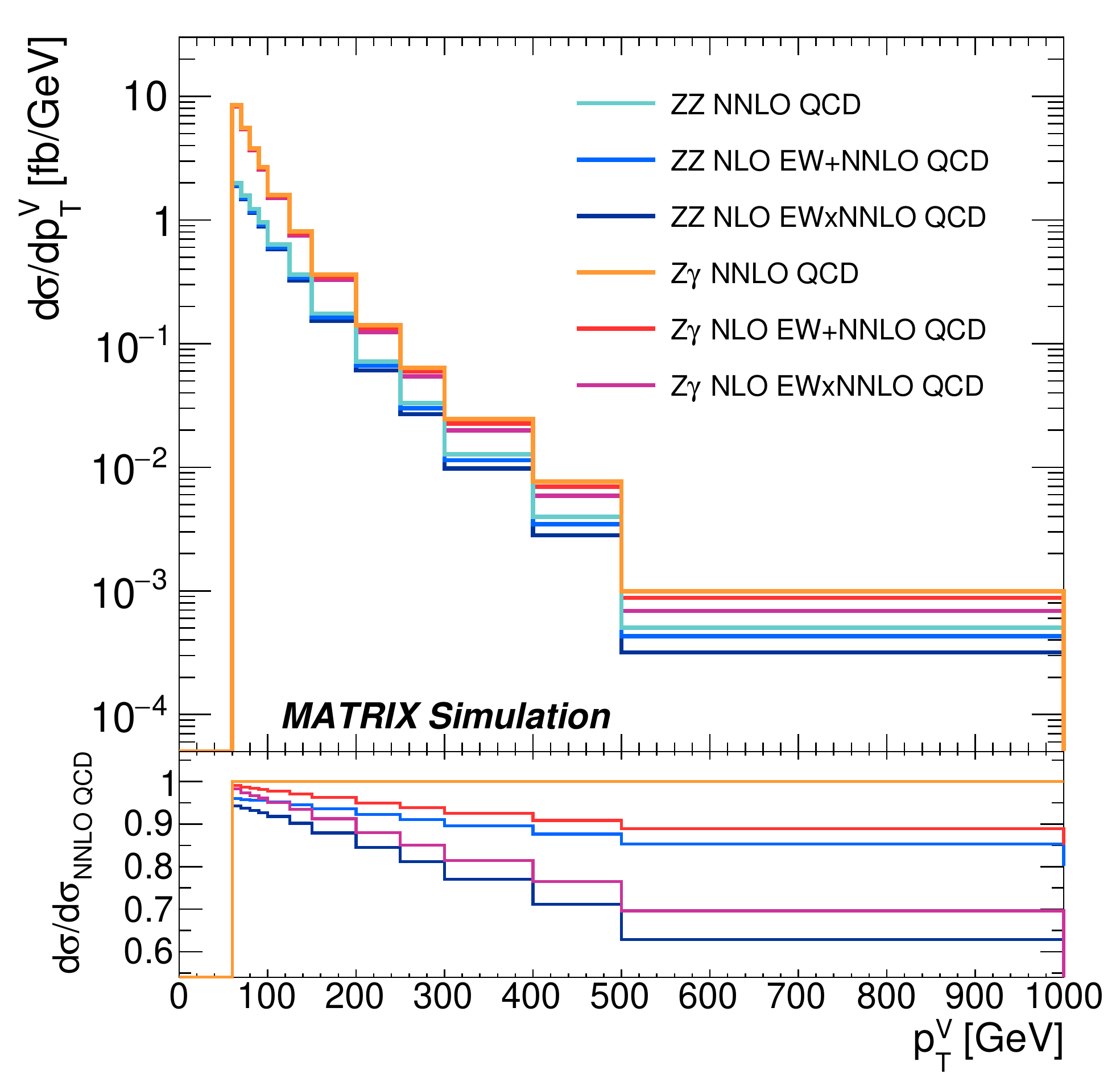}
\caption{}
\end{subfigure}
\hfill
\begin{subfigure}[b]{0.49\textwidth}
\includegraphics[width=\textwidth]{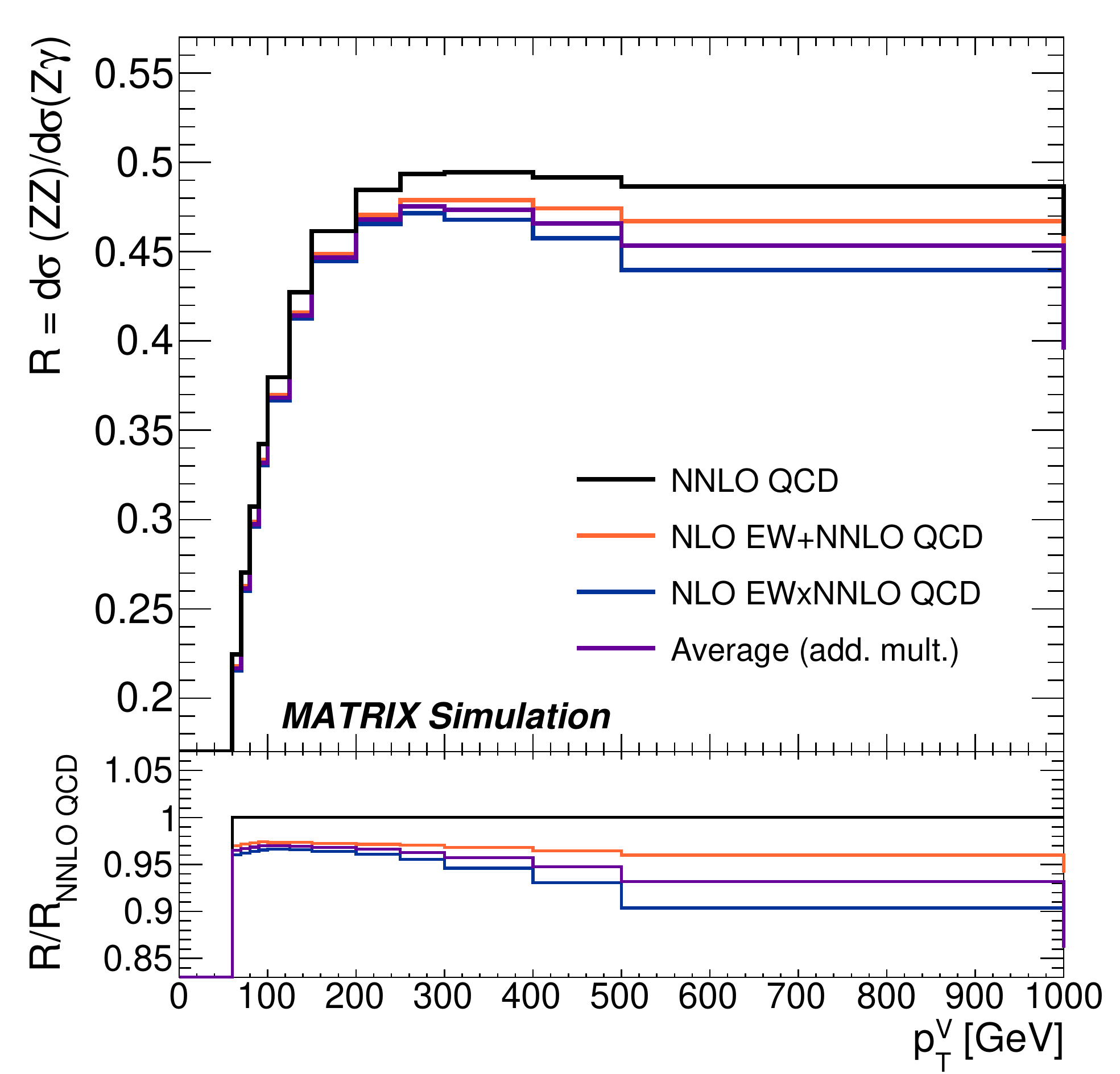}
\caption{}
\end{subfigure}
\caption{(a) NNLO QCD prediction and the combination of NNLO QCD and NLO EW calculations based on the multiplicative and additive prescriptions for the $ZZ(\to \ell^+\ell^-\nu\bar{\nu})$ (blue hues) and $Z\gamma(\to \ell^+\ell^-\gamma)$ (red hues) cross-sections. The bottom panel shows the cross-sections normalised to the NNLO QCD calculation. (b) $R$ at NNLO in QCD (black) and the combination of NNLO QCD and NLO EW calculations based on the multiplicative (blue) and additive prescriptions (orange), as well as their average (violet). The bottom panel shows the different $R$ normalised to the NNLO QCD calculation. 
}
\label{fig:EW}
\end{figure}

The NLO EW corrections applied in this study were provided by the authors of Ref.~\cite{Kallweit:2017khh} based on the OpenLoops generator~\cite{Buccioni:2019sur}. They can be combined with higher-order QCD corrections following one of two schemes:
\begin{equation}
\label{eq:EWcorrections}
    \begin{split}
         \text{Additive:~~} d\sigma_{\text{NNLO QCD+EW}} & = d\sigma_{\text{LO}} \left(1 +\Delta_{\text{QCD}}+\Delta_{\text{EW}} \right)+d\sigma^{gg}, \\
        \text{Multiplicative:~~} d\sigma_{\text{NNLO QCD} \times \text{EW}} & = d\sigma_{\text{LO}} 
    \left(1+\Delta_{\text{EW}})(1+\Delta_{\text{QCD}} \right) +d\sigma^{gg},
    \end{split}
\end{equation}
where $\Delta_{\text{QCD}}$ and ${\Delta_{\text{EW}}}$ correspond to the relative QCD and EW corrections, respectively. The multiplicative scheme is generally considered to be superior, however it can overestimate the corrections, so in this study the average of the additive and multiplicative results is used as the central value. The assigned uncertainty is the difference of this average with respect to the two schemes, as recommended in Ref.~\cite{Kallweit:2019zez}. Fig.~\ref{fig:EW} (a) shows the impact of the EW corrections on the \ptV\ distributions for both the $\ZZllnunu$ and \Zg\ processes. The correction grows towards more negative values with increasing \ptV\ and is larger for the $ZZ$ process than the $Z\gamma$ process.
The effect on $R$ is shown in Fig.~\ref{fig:EW} (b) to be about $-3$\% at low \ptV, growing to $-7$\% for \ptV $>$ 500~\GeV.

\begin{figure}[ht!]
\centering
\begin{subfigure}[b]{0.6\textwidth}
\includegraphics[width=0.9\textwidth]{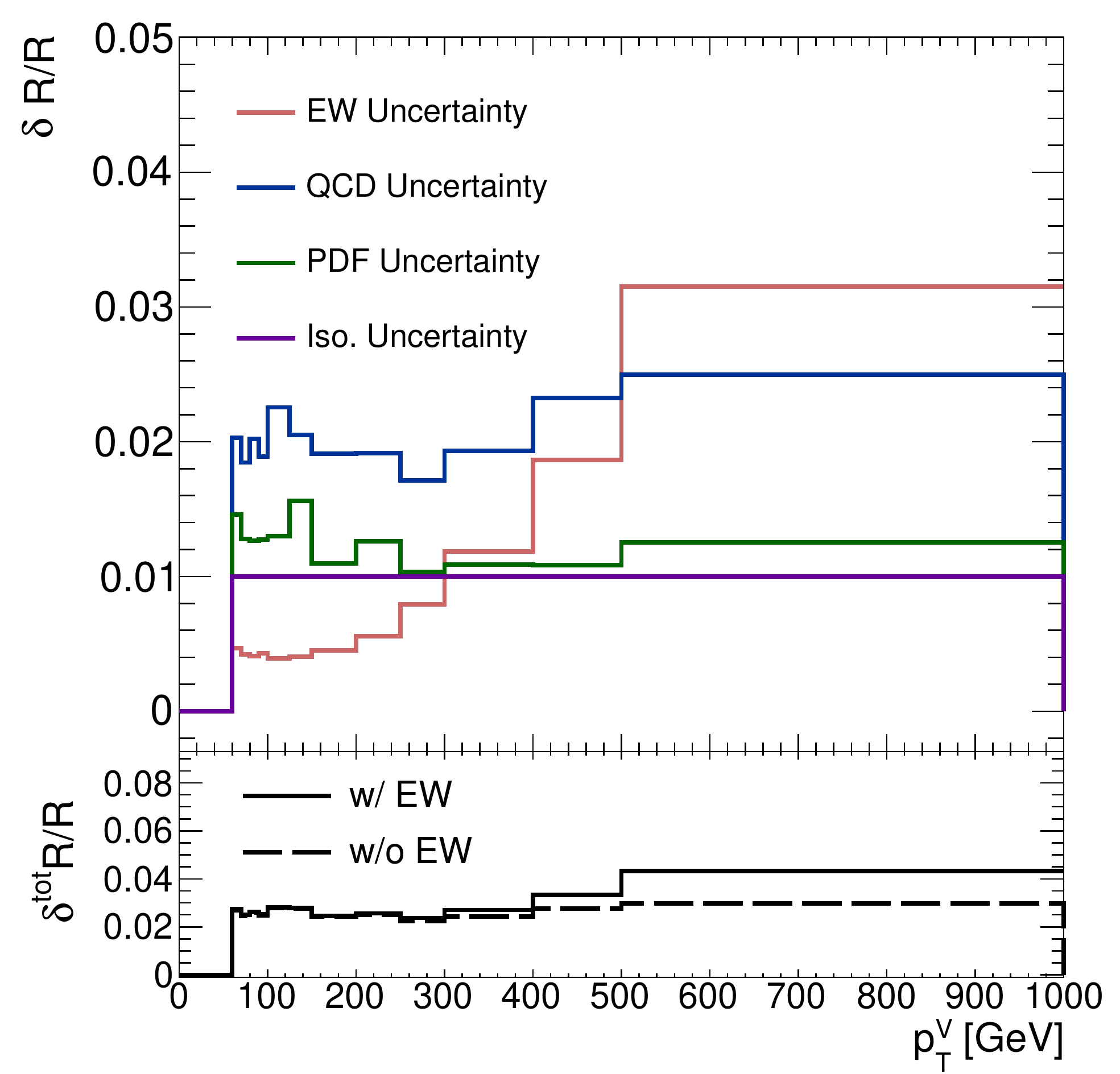}
\caption{}
\end{subfigure}
\begin{subfigure}[b]{0.6\textwidth}
\includegraphics[width=0.9\textwidth]{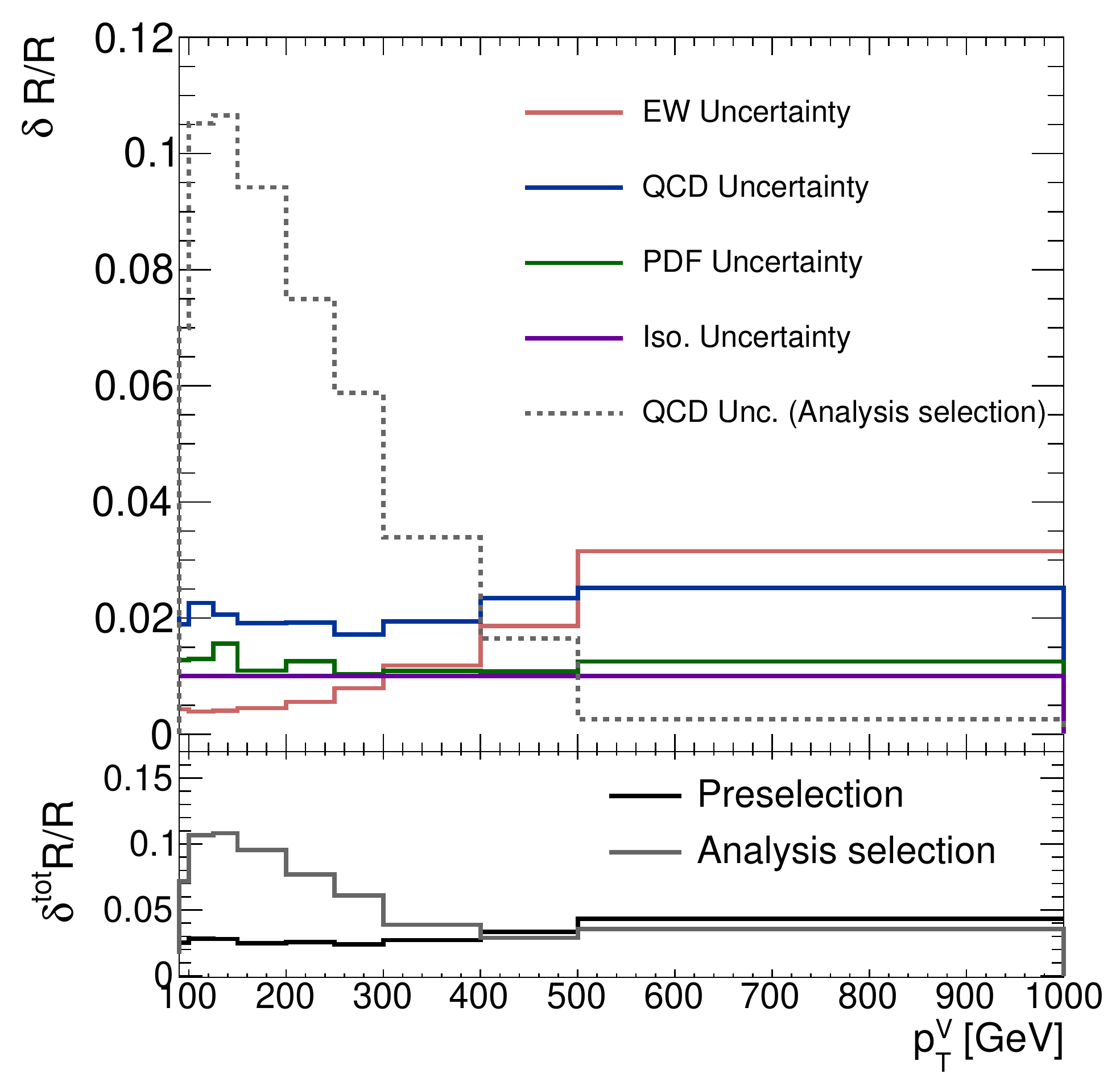}
\caption{}
\end{subfigure}
\caption{(a) Top frame: Relative uncertainties on $R$ after the baseline selection due to the EW corrections $\delta^{EW}$ (red), QCD effects $\delta^{QCD}$ (blue), PDF $\delta^{PDF}$ (green), and photon isolation $\delta^{iso}$ (violet). In the bottom frame the uncertainties are added in quadrature. The combined uncertainty is shown with (solid) and without (dashed) EW factorisation uncertainties.
(b) Same as (a), but including the QCD uncertainty when the additional selection from Table~\ref{tab:FullSRdef} is applied (dotted). This QCD uncertainty is only evaluated for \ptV $>$ 90~\GeV, as there are very few events below this value after the selection is applied.
The bottom frame compares the total uncertainty after the baseline selection (black) and after the additional selection (gray). 
}
\label{fig:combinedUnc}
\end{figure}

As a summary, Fig.~\ref{fig:combinedUnc} (a) shows the comparison of all determined uncertainty components on the cross-section ratio $R$ after the baseline selection. The bottom panel indicates the impact of the EW factorisation uncertainties on the total uncertainty. The EW factorisation uncertainty grows with \ptV\ and is the largest uncertainty ($\sim$3\%) for \ptV $>$ 500~\GeV. In Fig.~\ref{fig:combinedUnc} (b), the QCD uncertainties after the additional selection described in Table~\ref{tab:FullSRdef} are shown as well. The bottom panel compares  the total uncertainty after the baseline selection and after the additional selection. For \ptV $<$ 300 \GeV, the additional selection causes the QCD uncertainty to be as large as 10\%.

\FloatBarrier

\section{Conclusions}
In the context of searches for new phenomena at the ATLAS and CMS detectors in final states with leptons and large \MET, we explored the use of boson substitution for estimating the \ZZllnunu background from \Zg\ events measured in data. This article focuses on the cross-section ratio $R$ of the two processes and its uncertainties. $R$ is determined at NNLO in QCD, including NLO EW corrections, as a function of the transverse momentum of the substituted boson \ptV. Special care is taken to estimate the uncertainties due to the finite order of the QCD calculations  and due to photon isolation. EW uncertainties are included as well, based on the combination of QCD and EW higher-order corrections with different schemes. When selecting events with a $Z$ boson and missing transverse momentum or a photon, the resulting total uncertainties are $<3\%$ for 60~\GeV~$< \ptV  <$~400~\GeV, and grow to 4\% for higher \ptV. These uncertainties are generally smaller than the uncertainties on the individual processes, which is encouraging for experimental implementations of data-driven diboson estimates. However, if analysis selections include direct or indirect vetoes of additional QCD radiation, the QCD uncertainties on $R$ can increase to 10\% for \ptV $<$ 300~\GeV. 

\section*{Acknowledgments}
We wish to deeply thank M.~Grazzini for the many useful discussions, the help with running the \MATRIX software, and providing dedicated calculations for this work. We are also thankful to
S.~Kallweit and S.~Dittmaier for their calculations of the electroweak corrections. 
The work by V. Goumarre, S.~Heim, and B.~Heinemann was in part funded by the Deutsche Forschungsgemeinschaft under Germany‘s Excellence Strategy – EXC 2121 ``Quantum Universe" – 390833306. S. Heim thanks the Helmholtz Association for the support through the "Young Investigator Group" initiative.
This work has benefited from computing services provided by the German National Analysis Facility (NAF).

\section*{References}
\bibliographystyle{model1-num-names}
\bibliography{note.bib}

\clearpage
\appendix
\section{Isolation Cone}
\label{app:dyncone}

\begin{figure}[t]
\centering
\includegraphics[width=.7\textwidth]{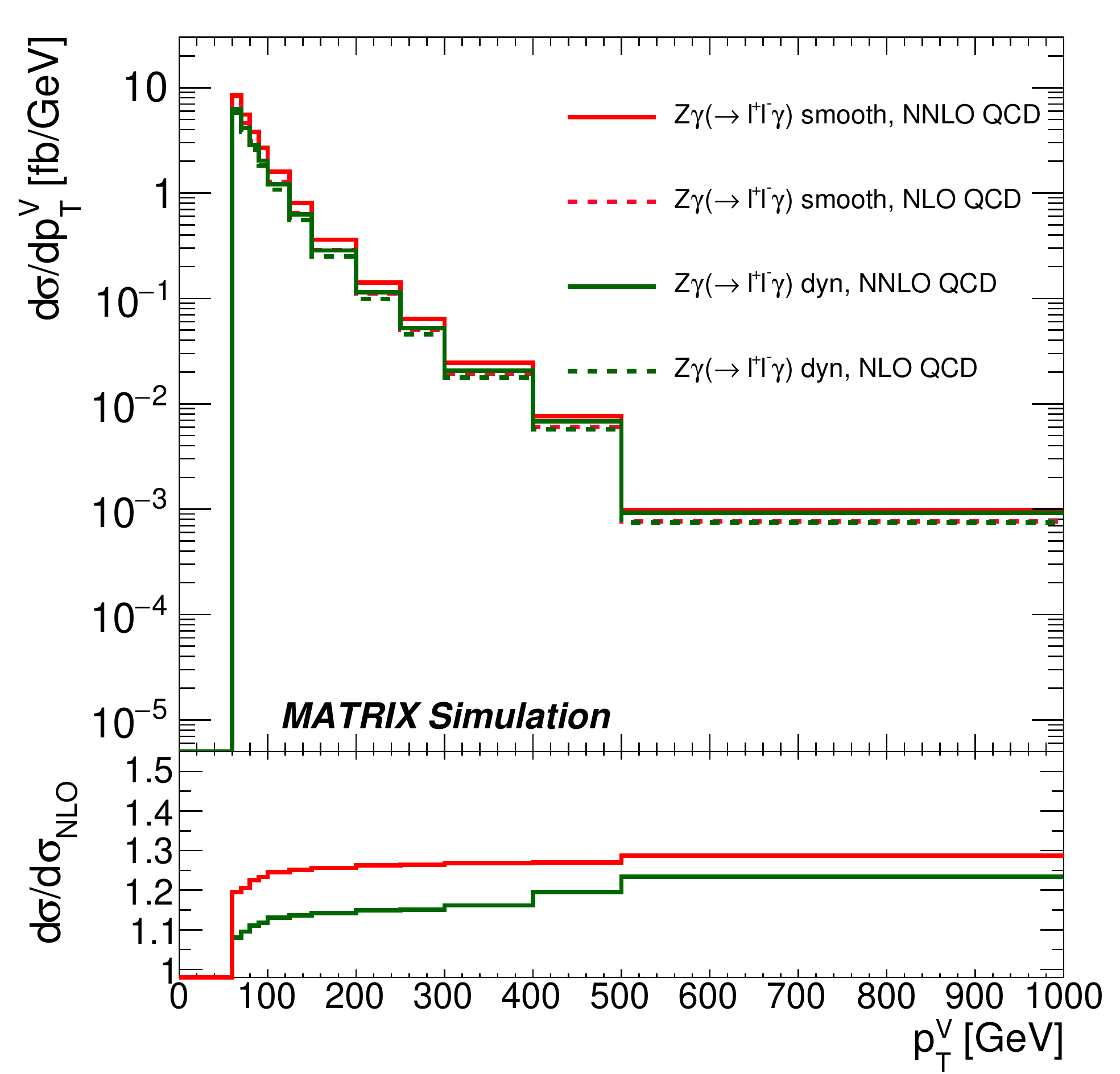}
\caption{\ptV distributions for $Z\gamma(\to \ell^+\ell^-\gamma)$ production at NLO (dashed) and NNLO (solid) in QCD, using the smooth cone isolation (red) and the dynamic cone isolation (green). The bottom frame shows the ratio of the NNLO calculation to the NLO prediction.}
\label{fig:Zgamma_dynXsec}
\end{figure}

It was already noted in Ref.~\cite{Lindert:2017olm} that the photon isolation requirement can alter the higher-order QCD corrections in the $Z\gamma$ calculation. This can lead to significantly different QCD corrections between the $Z\gamma$ and the $ZZ$ processes, as the latter does not require isolation. A supplementary isolation definition was proposed to account for this difference. The idea is to define the new isolation cone in such a way that both processes receive the same corrections at high $\ptV$. To this end, in Ref.~\cite{Lindert:2017olm}, a dynamic cone radius is defined
\begin{equation}
\label{eq:dynamic_cone}
    R_{dyn}(\pt^{\gamma},\epsilon_{\gamma}) = \frac{M_{Z}}{\pt^{\gamma}\sqrt{\epsilon_{\gamma}}},
\end{equation}
chosen such that the invariant mass of a collinear photon-jet pair ($R_{\gamma j} \ll 1$) is
\begin{equation}
    M^{2}_{\gamma j} \simeq M^{2}_{Z},
\end{equation}
whenever $\Delta R (\gamma,j) = R_{dyn}$ and $\pt^{j} = \epsilon_{\gamma}\pt^{\gamma}$. Replacing the radius $R_{0}$ in the smooth cone isolation definition in Eq.~\ref{eq:frixione_iso} with $R_{dyn}$ attempts to mimic the role of the $Z$ boson mass in the $ZZ$ process. It has to be noted that $R_{dyn}$ can become infinitely large at low $\pt^{\gamma}$, therefore a minimum radius $R_{min}$ has to be defined. In our study we choose $R_{min} = 1$, and the other two parameters to be the same as in the standard smooth cone isolation: $\varepsilon_{dyn} = 0.075, n_{dyn} = 1$. This means that the cone is only dynamic above $M_{Z}/\sqrt{\epsilon} \sim 330~\GeV$. Below this value, the dynamic cone isolation behaves effectively like the smooth cone isolation but with $R_{0} = 1$. In this region, the dynamic isolation is more restrictive than the smooth cone isolation, decreasing the cross-section. This can be observed in Fig.~\ref{fig:Zgamma_dynXsec}, where the NLO and NNLO cross-sections after the baseline selection are shown for both isolation prescriptions.

\begin{figure}[t]
\centering
\begin{subfigure}[b]{0.49\textwidth}
\includegraphics[width=\textwidth]{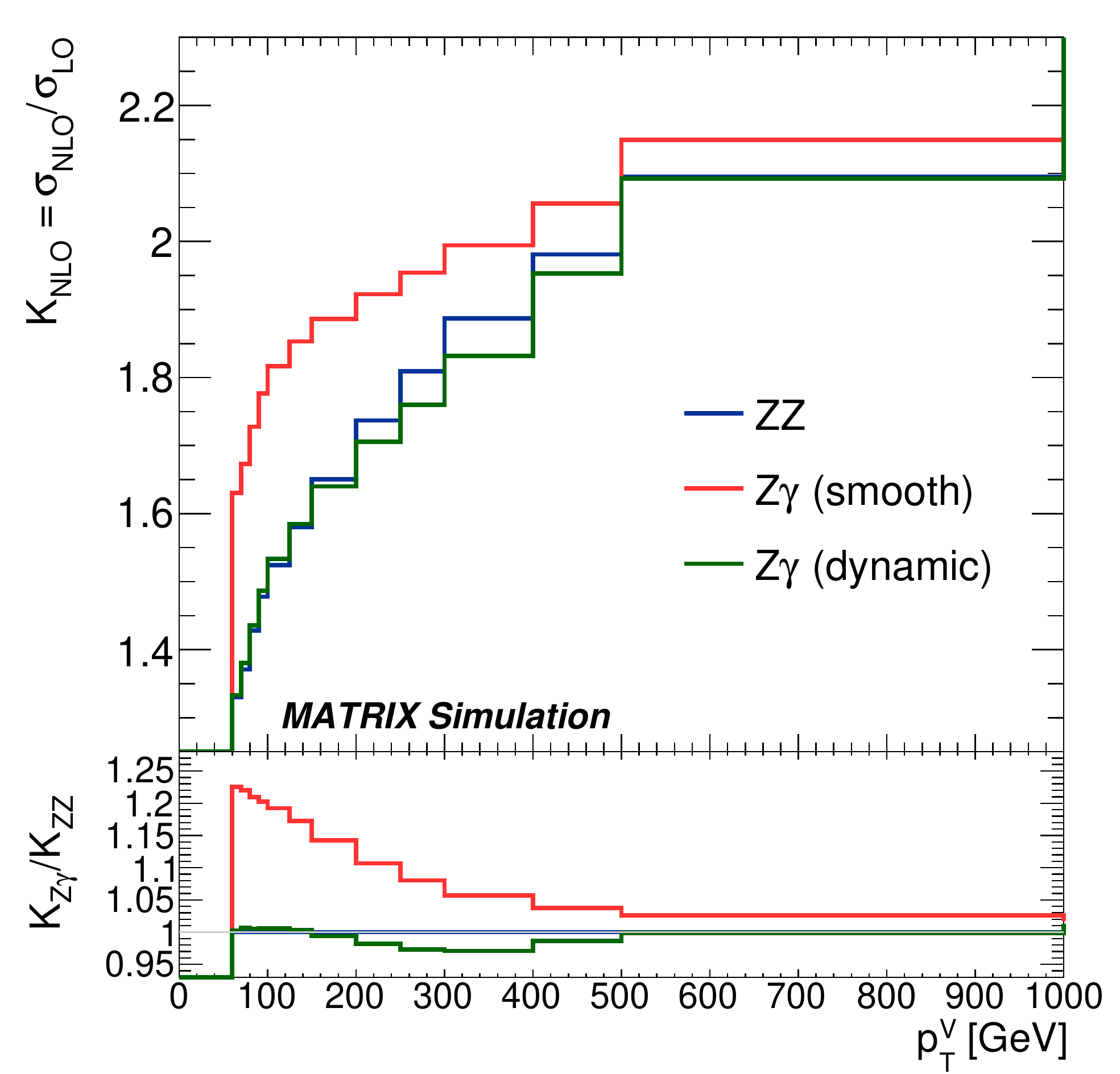}
\caption{}
\end{subfigure}
\begin{subfigure}[b]{0.49\textwidth}
\includegraphics[width=\textwidth]{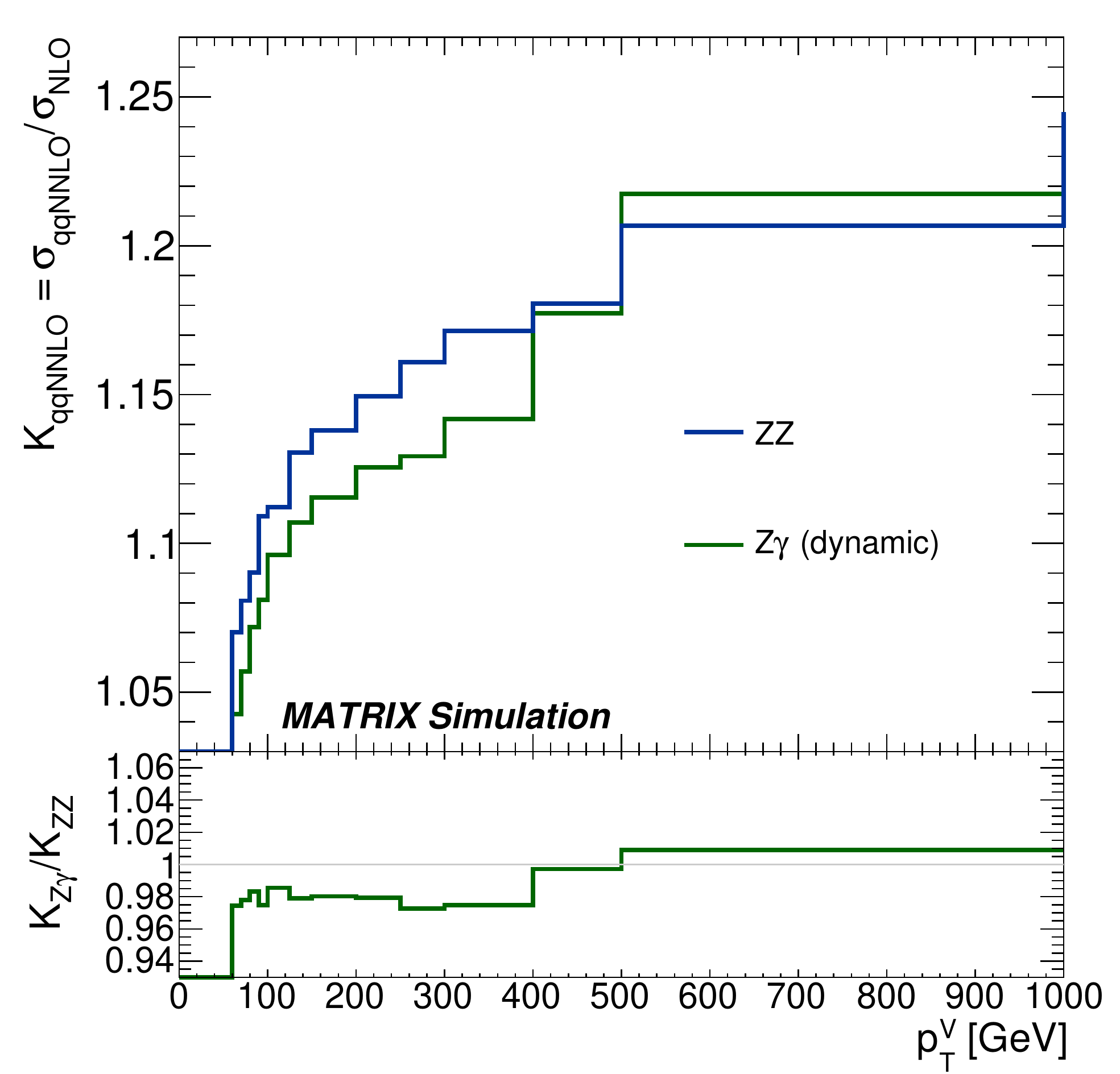}
\caption{}
\end{subfigure}
\begin{subfigure}[b]{0.49\textwidth}
\includegraphics[width=\textwidth]{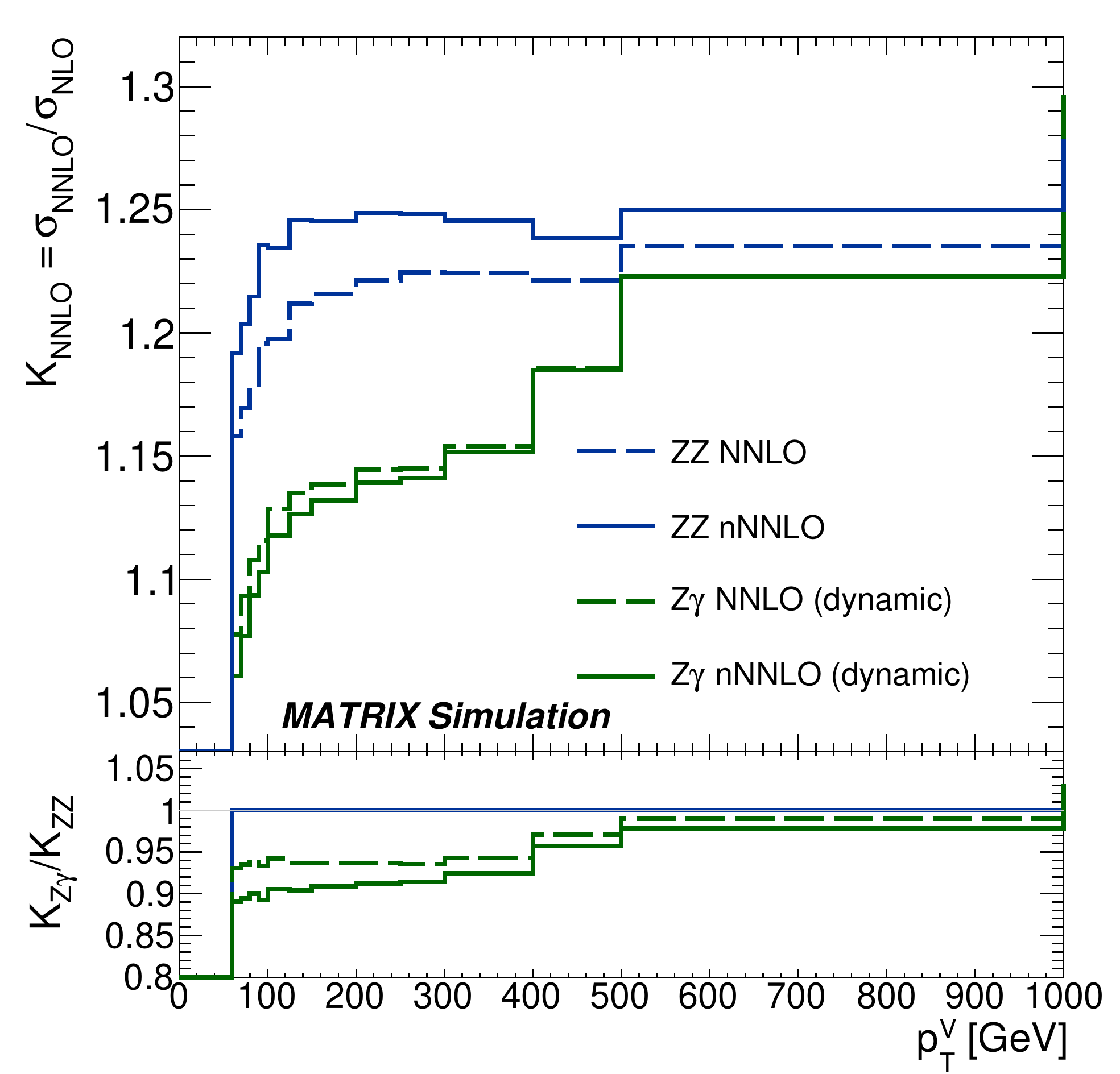}
\caption{}
\end{subfigure}
\caption{$K$-factors calculated at (a) NLO, (b) NNLO without including the gluon-induced processes (qqNNLO), and (c) NNLO and nNNLO in QCD for the $ZZ(\to \ell^+\ell^-\nu\bar{\nu})$ (blue) and $Z\gamma(\to \ell^+\ell^-\gamma)$ processes, the latter with smooth cone isolation (red) and with dynamic cone isolation (green) applied. The bottom panels show the ratio between the $Z\gamma$ and $ZZ$ $K$-factors.}
\label{fig:dynamic_cone}
\end{figure}

The NLO, qqNNLO, NNLO and nNNLO $K$-factors after the baseline selection are shown in Fig.~\ref{fig:dynamic_cone}. 
At NLO, especially at low $\ptV$, the agreement between the $ZZ$ and $Z\gamma$ $K$-factors is poor when applying the smooth cone isolation. There is a clear trend towards high \ptV, where the differences become smaller. On the other hand, when using the dynamic cone isolation, the differences stay below 3\% across the whole \ptV range. At qqNNLO, the dynamic cone isolation also keeps the difference between $ZZ$ and $Z\gamma$ $K$-factors below 3\%. Once the gluon-induced contribution is included (NNLO), the difference increases for \ptV $<$ 500~\GeV. Finally, at nNNLO the differences increases further to about 10\% at low \ptV, and below $5\%$ at high \ptV.

Fig.~\ref{fig:dynamic_cone} (a) in particular shows that for $\ptV > 330~\GeV$, the dynamic cone isolation changes the shape of the $K$-factor, whereas at low \ptV it simply causes a change in normalisation with respect to the smooth cone depending on the chosen parameter $R_{min}$. 

We decided not to include the dynamic cone isolation in this study since at NNLO or nNNLO it does not significantly reduce the QCD uncertainty compared to the smooth cone isolation, at least not for the current parameter choice. No further tuning of the parameters was attempted. A dynamic cone isolation would also require additional uncertainties due to the differences with respect to the isolation applied by the experiments.

\end{document}